\documentclass[final]{siamltex1213}
\usepackage{graphicx}
\usepackage{subfigure}
\usepackage{comment}
\usepackage{wrapfig}
\usepackage{amssymb}
\usepackage[cmex10]{amsmath}
\def\dt{\Delta t}
\graphicspath{{./figures/}}

\title{Semi-Implicit Time Integration of Atmospheric Flows with Characteristic-Based Flux Partitioning\thanks{This material is based upon work supported by the U.S. Department of Energy, Office of Science, Advanced Scientific Computing Research, under contract DE-AC02-06CH11357}} 

\author{Debojyoti Ghosh\footnotemark[2]\ \footnotemark[3]
\and Emil M. Constantinescu\footnotemark[2]\ \footnotemark[4]}

\begin{document}
\maketitle

\slugger{mms}{xxxx}{xx}{x}{x--x}

\renewcommand{\thefootnote}{\fnsymbol{footnote}}
\footnotetext[2]{Mathematics \& Computer Science Division, Argonne National Laboratory, Lemont, IL 60439}
\footnotetext[3]{\tt ghosh@mcs.anl.gov}
\footnotetext[4]{\tt emconsta@mcs.anl.gov}
\renewcommand{\thefootnote}{\arabic{footnote}}

\begin{abstract}
This paper presents a characteristic-based flux partitioning for the
semi-implicit time integration of atmospheric flows. Nonhydrostatic
models require the solution of the compressible Euler equations. The
acoustic time scale is significantly faster than the advective scale, yet
it is typically not relevant to atmospheric and weather phenomena. The
acoustic and advective components of the hyperbolic flux are separated
in the characteristic space. High-order, conservative additive
Runge-Kutta methods are applied to the partitioned equations so that
the acoustic component is integrated in time implicitly with an
unconditionally stable method, while the advective component is
integrated explicitly. The time step of the overall algorithm is thus
determined by the advective scale. Benchmark flow problems are used to
demonstrate the accuracy, stability, and convergence of the proposed
algorithm. The computational cost of the partitioned semi-implicit
approach is compared with that of explicit time integration. 
\end{abstract}



\pagestyle{myheadings}
\thispagestyle{plain}
\markboth{DEBOJYOTI GHOSH AND EMIL M. CONSTANTINESCU}{SEMI-IMPLICIT TIME INTEGRATION OF ATMOSPHERIC FLOWS}

\section{Introduction}
\label{sec:intro}

The simulation of mesoscale and limited-area atmospheric flows requires the solution to the compressible Euler equations, of which several formulations are used by operational weather prediction codes~\cite{giraldorestelli2008,giraldorestellilauter2010}. Expressing the governing equations in terms of the Exner pressure and potential temperature~\cite{gassmann,grell,coamps,janjic,arps} do not conserve mass, momentum, and energy. Alternatively, the equations are expressed as the conservation of mass, momentum, and potential temperature~\cite{ahmadlindeman,giraldokellyconsta2013,wrf,ullrich,yangcai2015} by assuming adiabatic flows~\cite{das}. Recent efforts~\cite{ahmad2007,botta2004,ghoshconstaAIAAJ2016,giraldorestelli2008,satoh} proposed solving the conservation laws for mass, momentum, and energy~\cite{laney}. If discretized by a conservative numerical method, this approach yields a truly conservative algorithm and allows for the specification of the true viscous terms. The Euler equations are characterized by two temporal scales---the acoustic and the advective scales.
Atmospheric flows are often low-Mach flows where the acoustic scale is significantly faster than
the advective scale~\cite{bourchtein}. The fluid velocities vary from 
stationary to $\sim30\,\textup{m}/\textup{s}$ within the troposphere
~\cite{wenschknothgalant}, resulting in Mach numbers lower than 
$\sim0.1$. In addition, the acoustic modes
do not affect weather phenomena significantly. 
  
Explicit time integration methods are inefficient because the largest stable
time step is restricted by the physically inconsequential
acoustic time scale. Implicit time integration methods can be
unconditionally stable; however, they have rarely been applied to
atmospheric flows~\cite{reisner2005,cyrneckels,yangcai2015}. One of their drawbacks is that they require the
solution of either a nonlinear system of equations or a linearized
approximation that introduces an error in the overall discretization.  
An alternative approach is an operator-split method, where the flux operator is split into its fast
(acoustic) and slow (advective) components and each component is
integrated in time separately. Split-explicit methods have been proposed
and applied to atmospheric flows~\cite{klempwilhelmson,wickerskamarock,klempskamarock,skamarockklemp2008,wicker,jebensknothweiner2009,wenschknothgalant}.
These methods are a form of decoupled 
multirate methods~\cite{Gear_1984,Constantinescu_A2007e,Sandu_A2009}. 

In this paper, we consider semi-implicit or implicit-explicit (IMEX) approaches that stabilize the
fast modes by integrating them implicitly in time; time-step sizes are
thus dictated by the slow scales. Semi-implicit methods for the
primitive meteorological equations were
introduced~\cite{kwizak,bourchtein} where the terms involving pressure
and gravitational forces are integrated implicitly.
A split-step semi-implicit method for the Euler equations expressed in terms of the
primitive flow variables was proposed~\cite{gattibono2006}; the prognostic variables
are perturbations to the hydrostatic mean profile, and the acoustic
modes are separated by decomposing the velocity into its anelastic, curl-free, and
harmonic components. Partially implicit peer methods were applied to the Euler
equations expressed in terms of the velocity and perturbations to the density
and potential temperature~\cite{jebensknothweiner2011}. Multistep IMEX methods based on the Adam's method and 
backward differencing were applied to the compressible Boussinesq equations~\cite{durranblossey,durran2010}. 
Other notable algorithms include a semi-Lagrangian 
semi-implicit method~\cite{bonaventura2000}, all-scale models~\cite{smolarkiewicz2014,benacchio},
and a split-step algorithm~\cite{wellershahrokhi}.
Drawbacks of these efforts include lack of
conservation (due to the form of the governing equations, the operator splitting for
semi-implicit time integration, or the choice of the implicit and explicit methods in the 
semi-implicit time integration) and lack of higher-than-second-order accuracy. An operator splitting was 
introduced for the governing equations expressed as perturbations to the
hydrostatic mean~\cite{giraldorestellilauter2010,giraldokellyconsta2013}; and integrated in time
by using multistep and multistage semi-implicit
methods to yield a conservative, high-order accurate algorithm.
In addition to scale separation between the acoustic and advective modes, splitting by dimension is
possible, leading to horizontally explicit, vertically implicit algorithms~\cite{satoh,ullrich,giraldokellyconsta2013}.

This paper presents a characteristic-based partitioning of the hyperbolic flux for the semi-implicit time integration of limited-area and mesoscale atmospheric flows. Our motivation is the development of a conservative, high-order accurate atmospheric flow solver based on the Euler equations expressed as the conservation of mass, momentum, and energy, with no other assumptions. The equations are not expressed as perturbations to a hydrostatic mean profile, and a well-balanced algorithm~\cite{ghoshconstaAIAAJ2016} is used to ensure numerical accuracy; we thus avoid any assumptions or manipulations specific to atmospheric flows. In contrast to previous approaches, we define the fast and slow components of the hyperbolic flux by partitioning it in the characteristic space. The discretized equations thus comprise scale-separated terms; eigenvalues of the fast term correspond to the acoustic mode, and the eigenvalues of the slow term correspond to the advective mode. In the context of implicit time integration methods, characteristic-based partitioning has been previously applied to selectively precondition the stiff characteristic modes of a hyperbolic system~\cite{reynoldswoodward}.  We linearize the partitioning such that the solution to a linear system of equations is required; in contrast, implicit time integration requires the solution to a nonlinear system of equations. Moreover, we show that this linearization does not introduce an error in the overall discretization. The partitioned equations are integrated in time with semi-implicit additive Runge-Kutta (ARK) methods~\cite{kennedycarpenter,giraldokellyconsta2013} implemented in the Portable, Extensible Toolkit for Scientific Computing (PETSc)~\cite{petsc-user-ref, petsc-web-page}.
We show that this partitioning of the flux allows time step sizes determined by the advective speeds. We also verify that the overall algorithm is conservative and achieves its theoretical orders of convergence.
Although atmospheric flows are low-speed flows, they often develop strong gradients, and stabilizing mechanisms are required~\cite{ahmadlindeman,giraldorestelli2008,ullrich,marrasnazarovgiraldo2015}.
In this paper, we use the fifth-order weighted essentially nonoscillatory (WENO)~\cite{liuosherchan,jiangshu} and the compact-reconstruction WENO (CRWENO)~\cite{ghoshbaederSISC2012,ghoshbaederJSC2014,ghoshmedidabaederAIAA2014} schemes for the spatial discretization. The algorithm described here is implemented in HyPar~\cite{hypar}, an open-source conservative finite-difference solver for hyperbolic-parabolic partial differential equations (PDEs).

The paper is organized as follows. Section~\ref{sec:govern} describes the governing equations, and Section~\ref{sec:nummeth} outlines the overall numerical method, including the spatial discretization. The characteristic-based flux partitioning is introduced in Section~\ref{sec:cbfp}. Section~\ref{sec:time} describes the semi-implicit time integration and the implementation of the linearized characteristic-based partitioning with multistage ARK methods. The extension to two-dimensional flows is presented in Section~\ref{sec:2d}. The proposed algorithm is tested for small problems in Section~\ref{sec:numtests} and applied to atmospheric flow problems in Section~\ref{sec:atmosflows}. Section~\ref{sec:conc} contains concluding remarks.

\section{Governing Equations}
\label{sec:govern}

The governing equations for limited-area and mesoscale nonhydrostatic atmospheric flows are the
Euler equations~\cite{laney}, with the addition of gravitational force as a source term.  They are expressed as
\begin{eqnarray}
\frac{\partial \rho}{\partial t} + \nabla \cdot \left( \rho {\bf u} \right) &=& 0, \label{eqn:mass_cons} \\
\frac{\partial \left(\rho {\bf u}\right)}{\partial t} + \nabla \cdot \left( \rho {\bf u} \otimes {\bf u} + p\mathcal{I}_d \right) &=& - \rho {\bf g}, \label{eqn:mom_cons}\\
\frac{\partial e}{\partial t} + \nabla \cdot \left( e+p \right){\bf u} &=& -\rho {\bf g} \cdot {\bf u}, \label{eqn:energy_cons}
\end{eqnarray}
where $\rho$ is the density, ${\bf u}$ is the velocity vector, $p$ is
the pressure, and ${\bf g}$ is the gravitational force vector (per
unit mass). $\mathcal{I}_d$ denotes the identity matrix of size $d$,
where $d$ is the number of space dimensions, and $\otimes$ represents
the Kronecker product. The energy is given by
\begin{equation}
e = \frac{p}{\gamma-1} + \frac{1}{2} \rho {\bf u} \cdot {\bf u},
\end{equation}
where $\gamma = 1.4$ is the specific heat ratio. The equation of state relates the pressure, density, and temperature as $p = \rho R T$, where $R$ is the universal gas constant and $T$ is the temperature. Two additional quantities of interest in atmospheric flows are the Exner pressure $\pi$ and the potential temperature $\theta$, defined as
\begin{equation}
\pi = \left(\frac{p}{p_0} \right)^{\frac{\gamma-1}{\gamma}}\ {\rm and}\ \theta = \frac{T}{\pi},
\end{equation}
respectively. The pressure at a reference altitude is denoted by $p_0$. We consider one- and two-dimensional flows ($d=1,2$) in this paper. The governing equations share the same form as (\ref{eqn:mass_cons})--(\ref{eqn:energy_cons}) when expressed in terms of nondimensional variables~\cite{ghoshconstaAIAAJ2016}, and thus these equations are used for both dimensional and nondimensional problems.

\section{Numerical Methodology}
\label{sec:nummeth}

\begin{figure}[t!]
\begin{center}
\includegraphics[width=0.75\textwidth]{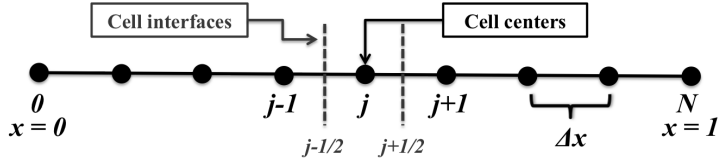}
\caption{Illustration of a one-dimensional domain and the grid on which (\ref{eqn:hyp_cons}) is discretized.}
\label{fig:domain1D_disc}
\end{center}
\end{figure}

The numerical discretization of the governing equations is described
in one spatial dimension, and it can be trivially extended to multiple dimensions. Equations~(\ref{eqn:mass_cons})--(\ref{eqn:energy_cons}) (with $d=1$) can be expressed as a system of hyperbolic PDEs,
\begin{equation}\label{eqn:hyp_cons}
\frac{\partial {\bf q}}{\partial t} + \frac{\partial {\bf f}\left({\bf q}\right)}{\partial x} = {\bf s}\left({\bf q}\right),
\end{equation}
where
\begin{equation}
{\bf q} = \left[\begin{array}{c} \rho \\ \rho u \\ e \end{array}\right],
{\bf f} = \left[\begin{array}{c} \rho u \\ \rho u^2 + p \\ (e+p) u \end{array}\right],
\ {\rm and}\ 
{\bf s} = \left[\begin{array}{c} 0 \\ - \rho g \\ -\rho u g \end{array}\right].
\end{equation}
Equation~(\ref{eqn:hyp_cons}) is discretized in space with a conservative finite-difference formulation. Figure~\ref{fig:domain1D_disc} shows a one-dimensional domain of unit length, discretized by $N+1$ grid points. The cell centers and interfaces are shown. The resulting semi-discrete ODE in time is given by
\begin{equation}\label{eqn:semidisc_ode}
\frac{d {\bf Q}}{dt} = \hat{\bf F}\left({\bf Q}\right) + \hat{\bf S}\left({\bf Q}\right),
\end{equation}
where ${\bf Q} = \left[{\bf q}_j; j = 1,\cdots,N-1 \right]$ is the solution vector of the state variable at the cell centers (excluding boundary points), $\hat{\bf S}$ is the discretized source term, and the discretized hyperbolic flux at a grid point is given by
\begin{equation}\label{eqn:cons_fd}
\hat{\bf F}_j = - \frac{1}{\Delta x} \left( \hat{\bf f}_{j+1/2} - \hat{\bf f}_{j-1/2} \right).
\end{equation}
The numerical flux $\hat{\bf f}$ is an approximation to the primitive
of ${\bf f}\left({\bf q}\right)$ at the cell interfaces $x_{j\pm 1/2}$.

Equation (\ref{eqn:hyp_cons}) represents a hyperbolic balance law
that admits equilibrium states where the pressure gradient is balanced
by the gravitational force. The spatially discretized ODE,
(\ref{eqn:semidisc_ode}), must preserve this balance on a finite grid
to machine precision; failure to do so will result in inaccurate
solutions since atmospheric phenomena are often small perturbations
around this balanced equilibrium state. We use a well-balanced formulation
to evaluate the source term $\hat{\bf S}$~\cite{ghoshconstaAIAAJ2016}. The 
description of this is omitted because it is independent of the time integration aspects discussed here; however, it is a necessary component of the overall algorithm. 

The numerical flux at the cell interfaces $\hat{\bf f}_{j \pm 1/2}$ in (\ref{eqn:cons_fd}) is computed by using the Rusanov upwinding scheme~\cite{rusanov,leveque},
\begin{equation}\label{eqn:rusanov}
\hat{\bf f}_{j+1/2} = \frac{1}{2} \left[ \hat{\bf f}^L_{j+1/2} + \hat{\bf f}^R_{j+1/2} - \left(\max_{j,j+1} \nu\right) \left( \hat{\bf q}^R_{j+1/2} - \hat{\bf q}^L_{j+1/2} \right) \right],
\end{equation}
where the superscripts $L$ and $R$ indicate the left- and right-biased interpolations, respectively. The dissipation factor is $\nu = a + \left|u\right|$, where $a = \sqrt{\gamma p/\rho}$ is the speed of sound. The left- and right-biased flux $\hat{\bf f}_{j+1/2}^{L,R}$ and solution $\hat{\bf q}^{L,R}_{j+1/2}$ at the interfaces are computed by using the fifth order WENO~\cite{jiangshu} and CRWENO~\cite{ghoshbaederSISC2012} schemes. The following paragraphs describe a left-biased reconstruction; the corresponding expressions for the right-biased reconstruction can be trivially obtained.
The description below applies to a scalar flux
function, and it is extended to the vector flux in \eqref{eqn:semidisc_ode} through a componentwise approach.

The WENO schemes use a solution-dependent interpolation stencil selection~\cite{liuosherchan} to achieve high-order accuracy where the solution is smooth and to avoid oscillations across discontinuities.
The fifth-order WENO scheme~\cite{jiangshu} is constructed by three third-order interpolation schemes:
\begin{eqnarray}
\hat{f}_{j+1/2}^1 &=& \frac{1}{3} f_{j-2} - \frac{7}{6} f_{j-1} + \frac{11}{6} f_j,\ c_1 = \frac{1}{10}, \label{eqn:third_order_cand1}\\
\hat{f}_{j+1/2}^2 &=& -\frac{1}{6} f_{j-1} + \frac{5}{6} f_j + \frac{1}{3} f_{j+1},\ c_2 = \frac{6}{10}, \label{eqn:third_order_cand2}\\
\hat{f}_{j+1/2}^3 &=& \frac{1}{3} f_j + \frac{5}{6} f_{j+1} - \frac{1}{6} f_{j+2},\ c_3 = \frac{3}{10}. \label{eqn:third_order_cand3}
\end{eqnarray}
Multiplying (\ref{eqn:third_order_cand1})--(\ref{eqn:third_order_cand3}) with their optimal coefficient $c_k,\ k=1,2,3$, and then adding them results in the following fifth-order accurate interpolation scheme:
\begin{equation}\label{eqn:fifth_order}
\hat{f}_{j+1/2} = \frac{1}{30} f_{j-2} - \frac{13}{60} f_{j-1} + \frac{47}{60} f_j + \frac{27}{60} f_{j+1} - \frac{1}{20} f_{j+2}.
\end{equation}
Solution-dependent weights are computed based on the local solution smoothness as
\begin{equation}\label{eqn:weno_weights}
\omega_k = \frac{\alpha_k}{\sum_k \alpha_k};\ \alpha_k = \frac{c_k}{\left(\epsilon + \beta_k\right)^p};\ k = 1,2,3,
\end{equation}
where $\epsilon=10^{-6}$ is a small number to prevent division by zero and $\beta_k$ are the smoothness indicators for the stencils, given by
\begin{eqnarray}
\beta_1 &=& \frac{13}{12} (f_{j-2}-2f_{j-1}+f_{j})^2 + \frac{1}{4}(f_{j-2}-4f_{j-1}+3f_{j})^2, \label{eqn:weno5is1} \\
\beta_2 &=& \frac{13}{12} (f_{j-1}-2f_{j}+f_{j+1})^2 + \frac{1}{4}(f_{j-1}-f_{j+1})^2, \label{eqn:weno5is2} \\
{\rm and}\ \beta_3 &=& \frac{13}{12} (f_{j}-2f_{j+1}+f_{j+2})^2 + \frac{1}{4}(3f_{j}-4f_{j+1}+f_{j+2})^2. \label{eqn:weno5is3}
\end{eqnarray}
The fifth-order WENO (WENO5) scheme is obtained by multiplying (\ref{eqn:third_order_cand1})--(\ref{eqn:third_order_cand3}) by the solution-dependent weights $\omega_k$ (instead of the optimal coefficients $c_k$) and then adding them. It can be expressed as
\begin{align}\label{eqn:weno5}
\hat{f}_{j+1/2} =& \frac{\omega_1}{3} f_{j-2} - \frac{1}{6}(7\omega_1+\omega_2)f_{j-1} + \frac{1}{6}(11\omega_1+5\omega_2+2\omega_3)f_j\nonumber \\ 
&+ \frac{1}{6}(2\omega_2+5\omega_3)f_{j+1} - \frac{\omega_3}{6}f_{j+2}.
\end{align}
If the solution is locally smooth, $\omega_k \rightarrow c_k$, $k=1,2,3$, and (\ref{eqn:weno5}) is equivalent to (\ref{eqn:fifth_order}). 

The CRWENO scheme~\cite{ghoshbaederSISC2012} applies the WENO concept of solution-dependent interpolation stencils to compact finite-difference methods~\cite{lele}. The fifth-order CRWENO scheme~\cite{ghoshbaederSISC2012,ghoshbaederJSC2014} is constructed by considering three third-order compact interpolation schemes:
\begin{eqnarray}
\frac{2}{3}\hat{f}_{j-1/2} + \frac{1}{3}\hat{f}_{j+1/2} = \frac{1}{6} \left( f_{j-1} + 5f_j \right);&\ &c_1 = \frac{2}{10}, \label{eqn:s1} \\
\frac{1}{3}\hat{f}_{j-1/2}+\frac{2}{3}\hat{f}_{j+1/2} = \frac{1}{6} \left( 5f_j + f_{j+1} \right);&\ &c_2 = \frac{5}{10}, \label{eqn:s2}\\
\frac{2}{3}\hat{f}_{j+1/2} + \frac{1}{3}\hat{f}_{j+3/2} = \frac{1}{6} \left( f_j + 5f_{j+1} \right);&\ &c_3 = \frac{3}{10}. \label{eqn:s3}
\end{eqnarray}
Multiplying (\ref{eqn:s1})--(\ref{eqn:s3}) with their optimal coefficients ($c_k,\ k=1,2,3$) and adding them results in a fifth-order compact scheme:
\begin{equation}\label{eqn:compact_5thorder}
\frac{3}{10}\hat{f}_{j-1/2} + \frac{6}{10}\hat{f}_{j+1/2} + \frac{1}{10}\hat{f}_{j+3/2} = \frac{1}{30}f_{j-1} + \frac{19}{30}f_j + \frac{1}{3}f_{j+1}.
\end{equation}
Replacing the optimal coefficients $c_k$ with solution-dependent weights $\omega_k$ yields the fifth-order CRWENO scheme (CRWENO5):
\begin{eqnarray}
\left(\frac{2}{3}\omega_1+\frac{1}{3}\omega_2\right)\hat{f}_{j-1/2} & + & \left[\frac{1}{3}\omega_1+\frac{2}{3}(\omega_2+\omega_3)\right]\hat{f}_{j+1/2} + \frac{1}{3}\omega_3\hat{f}_{j+3/2} \nonumber \\ & = & \frac{\omega_1}{6}f_{j-1} + \frac{5(\omega_1+\omega_2)+\omega_3}{6}f_j + \frac{\omega_2+5\omega_3}{6}f_{j+1} \label{eqn:crweno5}.
\end{eqnarray}
The weights $\omega_k$ are computed by (\ref{eqn:weno_weights}) and (\ref{eqn:weno5is1})--(\ref{eqn:weno5is3}). If the solution is locally smooth, $\omega_k \rightarrow c_k$, $k=1,2,3$, and (\ref{eqn:crweno5}) is equivalent to (\ref{eqn:compact_5thorder}). The left-hand side of (\ref{eqn:crweno5}) represents a tridiagonal system with solution-dependent coefficients that needs to be solved at each time-integration step or stage. An efficient and scalable implementation of the CRWENO5 scheme~\cite{ghoshconstbrownSISC2015} is used in this study.

\section{Characteristic-Based Flux Partitioning}
\label{sec:cbfp}

\begin{figure}[t!]
\begin{center}
\subfigure[Spatial discretization $\mathcal{D}$]                                    {\includegraphics[width=0.32\textwidth]{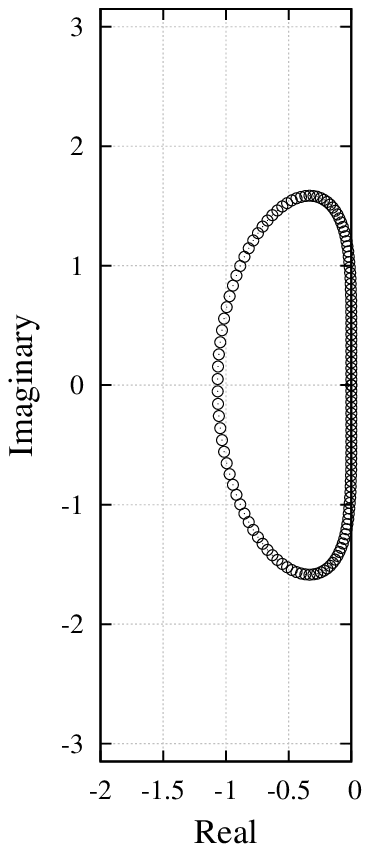}\label{fig:eigenvalues_weno5}}
\subfigure[Right-hand-side operator $\frac{\partial\hat{\bf F}}{\partial {\bf Q}}$] {\includegraphics[width=0.32\textwidth]{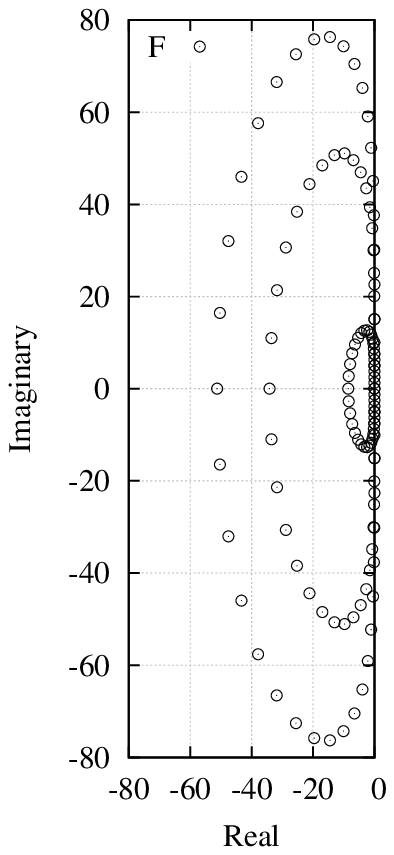}  \label{fig:eigenvalues_rhs}}
\subfigure[Split operators $\frac{\partial\hat{\bf F}_{F,S}}{\partial {\bf Q}}$]    {\includegraphics[width=0.32\textwidth]{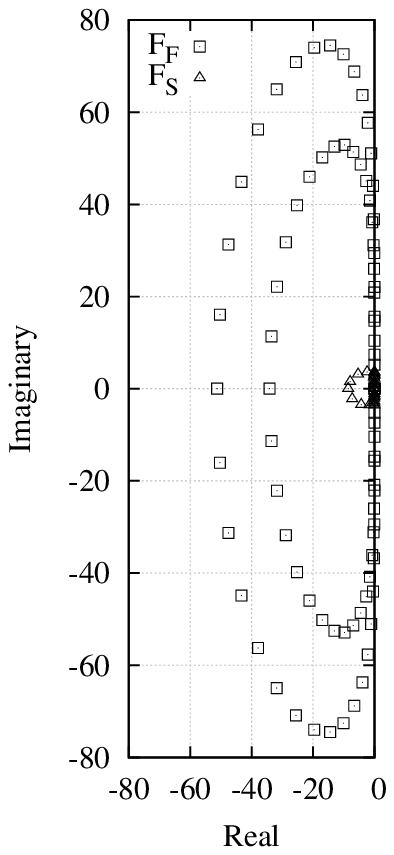}\label{fig:eigenvalues_split}}
\end{center}
\caption{Eigenvalues of the spatial discretization operator corresponding to WENO5, the Jacobian of the right-hand side of (\ref{eqn:semidisc_ode2}), and the Jacobians of the fast and slow partitioned terms of (\ref{eqn:semidisc_ode_part}). Note that the eigenvalues shown in (b) are those shown in (a) scaled by $\left\{u,u\pm a\right\}/\Delta x$.}
\label{fig:eigenvalues}
\end{figure}

The separation of the acoustic and advective components of the hyperbolic flux is described by considering (\ref{eqn:hyp_cons}) and its semi-discrete form (\ref{eqn:semidisc_ode}), without the source terms. The one-dimensional Euler equations, although nonlinear, satisfy the following property~\cite{laney},
\begin{equation}
  {\bf f}\left({\bf q}\right) =
  \mathcal{A}\left({\bf q}\right){\bf q},\mathcal{A}\left({\bf
    q}\right) =
  \frac{\partial{\bf f}}{\partial{\bf q}},
\end{equation}
where $\mathcal{A}$ is the flux Jacobian. This property, though not essential to the flux partitioning, is useful as a tool to describe it. Equation (\ref{eqn:semidisc_ode}) (without the source term) can be expressed as
\begin{equation}\label{eqn:semidisc_ode2}
\frac{d {\bf Q}}{dt} = \hat{\bf F} \left({\bf Q}\right) \equiv \left[ \mathcal{D} \otimes \mathcal{A} \right] {\bf Q},
\end{equation}
where $\mathcal{D}$ represents a finite-difference operator for a scalar function $\phi\left(x\right)$ on a grid,
\begin{equation}
-\left[\phi_{x,j} \right] = \left[\mathcal{D}\right] \left[ \phi_j \right] + O\left(\Delta x^r\right),\ 0 < j < N,\ \phi_j = \phi\left(x_j\right),\ \phi_{x,j} = \phi_x\left(x_j\right)
\end{equation}
with $r$ being the spatial order of accuracy. The WENO5 and CRWENO5 schemes, described in the preceding section, can be expressed in this form~\cite{ghoshPhDthesis}. Therefore, the eigenvalues of the right-hand side (RHS) operator of (\ref{eqn:semidisc_ode2}) are the products of the eigenvalues of the discretization operator $\mathcal{D}$ and the eigenvalues of the flux Jacobian that are the characteristic wave speeds of the Euler equations,
\begin{equation}
\lambda\left(\frac{\partial \hat{\bf F}}{\partial {\bf Q}}\right) = \lambda\left(\mathcal{D}\right) \ast \lambda\left(\mathcal{A}\right),
\end{equation}
where $\ast$ denotes the following operation between two sets $A$ and $B$:
\begin{equation}
A \ast B = \left\{ \left(ab\right) | a \in A, b \in B \right\}.
\end{equation}
The flux Jacobian has three real eigenvalues~\cite{laney},
\begin{equation}
\lambda\left(\mathcal{A}\right) = \left\{ u, u+a, u-a \right\},
\end{equation}
where $u$ is the flow velocity and $a$ is the local speed of sound. Figure~\ref{fig:eigenvalues_weno5} shows the eigenvalues of the finite-difference operator $\mathcal{D}$ representing the WENO5 scheme, computed by using a linear spectral analysis~\cite{ghoshbaederJSC2014}. Figure~\ref{fig:eigenvalues_rhs} shows the eigenvalues of the Jacobian of $\hat{\bf F}$ evaluated on a periodic domain of unit length, discretized by a grid with $40$ points and the WENO5 scheme, with $\rho = 1 + 0.1\sin\left(2\pi x\right)$, $u = 0.2$, $p = 1/\gamma$. The mean speed of sound is $a_\infty = \sqrt{\gamma p_\infty/\rho_\infty} = 1$, and therefore the mean Mach number is $M_\infty = u_\infty/a_\infty = 0.2$. The Jacobian of $\hat{\bf F}$ is computed by using finite differences. The eigenvalues in Figure~\ref{fig:eigenvalues_rhs} form three distinct sets that correspond to the eigenvalues of $\mathcal{D}$ (in Figure~\ref{fig:eigenvalues_weno5}) multiplied by each of the characteristic wave speeds of the Euler equations. The smallest ring represents the advective mode ($u$) where the eigenvalues of $\mathcal{D}$ are scaled by $u/\Delta x$. The two larger rings represent the acoustic modes ($u\pm a$) where the eigenvalues of $\mathcal{D}$ are scaled by $\left(u \pm a\right)/\Delta x$. The separation in magnitude of the acoustic and advective eigenvalues is a function of the Mach number $M=u/a$; lower Mach numbers result in a larger separation.

The flux term ${\bf f}\left({\bf q}\right)$ is partitioned into its slow and fast components as follows:
\begin{equation}
{\bf f}\left({\bf q}\right) = \mathcal{A}\left({\bf q}\right){\bf q} = \mathcal{A}_F\left({\bf q}\right){\bf q} + \mathcal{A}_S\left({\bf q}\right){\bf q} = {\bf f}_F\left({\bf q}\right) + {\bf f}_S\left({\bf q}\right),
\end{equation}
where $\mathcal{A} = \mathcal{A}_F + \mathcal{A}_S$, and the subscripts $F$ and $S$ denote ``fast" and ``slow" time scales, respectively. The partitioned flux Jacobians $\mathcal{A}_{F,S}$ are defined as
\begin{equation}
\mathcal{A}_{F,S} = \mathcal{X}\Lambda_{F,S}\mathcal{X}^{-1} \label{eqn:fastslowjac},\ 
\Lambda_F = \left[\begin{array}{ccc} 0 & & \\ & u+a & \\ & & u-a \end{array}\right],\ 
\Lambda_S = \left[\begin{array}{ccc} u & & \\ & 0 & \\ & & 0 \end{array}\right], 
\end{equation}
where $\mathcal{X}$ is the matrix with the right eigenvectors as its columns
and $\mathcal{X}^{-1}$ is the matrix with the left eigenvectors as its rows.
$\Lambda_{F,S}$ represent the fast (acoustic) and slow (advective) characteristic modes and satisfy
\begin{align}
\Lambda_F &+ \Lambda_S = {\rm diag}\left[ u, u-a, u+a \right] = \mathcal{X}^{-1}\mathcal{A}\mathcal{X}.
\end{align}
The flux Jacobian $\mathcal{A}$, and the matrices $\mathcal{X},\mathcal{X}^{-1}$ for the one-dimensional Euler equations are provided in~\cite{laney}, 
and the resulting expressions for the slow and fast flux ${\bf f}_{S,F}\left({\bf q}\right) = \mathcal{A}_{S,F}{\bf q}$ are
\begin{equation}
{\bf f}_S\left({\bf q}\right) = \left[\begin{array}{c} \left(\frac{\gamma-1}{\gamma}\right)\rho u \\ \left(\frac{\gamma-1}{\gamma}\right) \rho u^2 \\ \frac{1}{2}\left(\frac{\gamma-1}{\gamma}\right)\rho u^3 \end{array}\right],
{\bf f}_F\left({\bf q}\right) = \left[\begin{array}{c} \left(\frac{1}{\gamma}\right)\rho u \\ \left(\frac{1}{\gamma}\right) \rho u^2 + p \\ \left(e+p\right)u - \frac{1}{2}\left(\frac{\gamma-1}{\gamma}\right)\rho u^3 \end{array}\right].
\end{equation}
The corresponding partitioning for the RHS operator $\hat{\bf F}$ of (\ref{eqn:semidisc_ode2}) is expressed as follows:
\begin{align}
\hat{\bf F}\left({\bf Q}\right) &= \left[ \mathcal{D} \otimes \mathcal{A} \right] {\bf Q} = \left[ \mathcal{D} \otimes \left( \mathcal{A}_F +  \mathcal{A}_S \right)\right] {\bf Q} \nonumber\\
&= \left[ \mathcal{D} \otimes \mathcal{A}_F \right]{\bf Q} + \left[ \mathcal{D} \otimes \mathcal{A}_S \right] {\bf Q} = \hat{\bf F}_F + \hat{\bf F}_S \label{eqn:fastslowpart},
\end{align}
where $\hat{\bf F}_{F,S}$ are the spatially discretized terms corresponding to the partitioned flux ${\bf f}_{F,S}$.
The fast term $\hat{\bf F}_F$ represents only the acoustic modes, while the slow term $\hat{\bf F}_S$ represents the advective mode. Figure~\ref{fig:eigenvalues_split} shows the eigenvalues of the Jacobians of the partitioned terms $\hat{\bf F}_{F,S}$ for the same flow as in Figure~\ref{fig:eigenvalues_rhs}. The partitioning results in a clear separation of the advective and acoustic eigenvalues; the eigenvalues of the slow term are significantly smaller in magnitude than those of the fast term. We note that the eigenvalues of the partitioned terms $\hat{\bf F}_{F,S}$ do not correspond exactly to the eigenvalues of $\hat{\bf F}$ because of the nonlinearity of the Euler equations,
\begin{equation}
\frac{\partial {\bf f}_{F,S}}{\partial q} \neq \mathcal{A}_{F,S} \Rightarrow \frac{\partial \hat{\bf F}_{F,S}}{\partial {\bf Q}} \neq \mathcal{D}\otimes\mathcal{A}_{F,S} \Rightarrow 
\Lambda\left[ \frac{\partial \hat{\bf F}_F}{\partial {\bf Q}} \right] \cup \Lambda\left[ \frac{\partial \hat{\bf F}_S}{\partial {\bf Q}} \right] \neq \Lambda\left[ \frac{\partial \hat{\bf F}}{\partial {\bf Q}} \right].
\end{equation}
Atmospheric flows are low-speed flows where the advective mode is significantly slower than the acoustic modes ($u \ll a$). The separation of the two time scales is useful in the context of semi-implicit time integration, discussed in the next section.

With the partitioning defined as (\ref{eqn:fastslowpart}), (\ref{eqn:semidisc_ode}) can be expressed as
\begin{equation}\label{eqn:semidisc_ode_part}
\frac{d {\bf Q}}{dt} = \left\{ \hat{\bf F}_F\left({\bf Q}\right) + \hat{\bf F}_S\left({\bf Q}\right) \right\} + \hat{\bf S}\left({\bf Q}\right).
\end{equation}
We note that (\ref{eqn:fastslowpart}) holds true if and only if both the slow and fast flux terms ${\bf f}_{F,S}$ are discretized by the same finite-difference operator $\mathcal{D}$. In the context of the nonlinear WENO5 and CRWENO5 schemes, this implies that the same solution-dependent coefficients for the interpolation operators (\ref{eqn:weno5}) or (\ref{eqn:crweno5}) need to be used for discretizing ${\bf f}_S$ and ${\bf f}_F$. In our implementation, the WENO coefficients (\ref{eqn:weno_weights}) are computed based on ${\bf f}\left({\bf q}\right)$. 

\section{Time Integration}
\label{sec:time}

Equation (\ref{eqn:semidisc_ode_part}) is integrated in time by using semi-implicit additive Runge-Kutta (ARK) methods~\cite{ascherruuthspiteri,kennedycarpenter,pareschirusso} implemented in the time integration module (TS) of PETSc~\cite{petsc-user-ref,petsc-web-page}. These methods apply two different integrators for the slow and the fast terms; the fast terms are integrated in time implicitly, and thus the largest stable time step size of the algorithm is determined by the eigenvalues of the slow term. ARK methods can be represented with the following Butcher tableaux~\cite{butcher2003}:
\begin{align}
\left(
\begin{array}{c|c}
  c_i & a_{ij}\\
\hline
  & b_j
\end{array},\ 
\begin{array}{c|c}
  \tilde{c}_i & \tilde{a}_{ij}\\
\hline
  & \tilde{b}_j
\end{array}; i,j = 1,\cdots,s
\right),\label{eqn:ark}
c_i = \sum_{j=1}^s a_{ij}, \tilde{c}_i = \sum_{j=1}^s \tilde{a}_{ij},
\end{align}
where $a_{ij},b_j,c_i$ define the explicit integrator for the slow
term, $\tilde{a}_{ij},\tilde{b}_j,\tilde{c}_i$ define the implicit
integrator for the fast term, and $s$ is the number of stages. The
coefficients satisfy $a_{ij} = 0, j \ge i$ and $\tilde{a}_{ij} = 0, j
> i$. The ARK methods applied to \eqref{eqn:semidisc_ode_part} and
using coefficients (\ref{eqn:ark}) result in the following:
\begin{subequations}
  \label{eq:ARK}
\begin{align}
{\rm Stage}\ &{\rm computations}\ i = 1,\cdots,s:\nonumber\\
{\bf Q}^{\left(i\right)} &= {\bf Q}^n + \Delta t \sum_{j=1}^{i-1} a_{ij} \hat{\bf F}_S\left({\bf Q}^{\left(j\right)}\right) + \Delta t \sum_{j=1}^{i} \tilde{a}_{ij} \left\{ \hat{\bf F}_F\left({\bf Q}^{\left(j\right)}\right) + \hat{\bf S}\left({\bf Q}^{\left(j\right)}\right) \right\},\label{eqn:stages}\\ 
{\rm Step}\ &{\rm completion}:\nonumber\\
{\bf Q}^{n+1} &= {\bf Q}^n + \Delta t \sum_{i=1}^s b_i \hat{\bf F}_S\left({\bf Q}^{\left(j\right)}\right) + \Delta t \sum_{i=1}^s \tilde{b}_i \left\{ \hat{\bf F}_F\left({\bf Q}^{\left(j\right)}\right) + \hat{\bf S}\left({\bf Q}^{\left(i\right)}\right)  \right\}, \label{eqn:stepcomp}
\end{align}
\end{subequations}
where ${\bf Q}^n$ is the solution at the current time step and ${\bf Q}^{n+1}$ is the solution at the next time step. The gravitational source term is treated implicitly in time.

\begin{table}[t!]
\begin{center}
\caption{List of time integration methods and their orders and number of stages.}
\label{tab:methods}
\begin{tabular*}{\textwidth}{@{\extracolsep{\fill}}| l | l | c c | l |}
\hline
Name & Type & Order & Stages ($s$) & Comments/Reference \\
\hline
ARK 2c & Semi-implicit & 2 & 3 & \cite{giraldokellyconsta2013} \\
ARK 3  & Semi-implicit & 3 & 4 & \cite{kennedycarpenter} \\
ARK 4  & Semi-implicit & 4 & 6 & \cite{kennedycarpenter} \\
RK  2a & Explicit      & 2 & 2 & Explicit midpoint method \\
RK  3  & Explicit      & 3 & 3 & Kutta's third-order method \\
RK  4  & Explicit      & 4 & 4 & Classical fourth-order method \\
\hline
\end{tabular*}
\end{center}
\end{table}

Three high-order ARK methods are considered in this study:
a second-order (three-stage) method (ARK 2c) constructed in
~\cite{giraldokellyconsta2013} and defined by
\begin{align}
\left(
\begin{array}{c|c c c}
  0 & 0 & & \\
  2-\sqrt{2} & 2-\sqrt{2} & 0 & \\
  1 & 1-a_{3,2} & a_{3,2} & 0 \\
\hline
  & \frac{1}{2\sqrt{2}} & \frac{1}{2\sqrt{2}} & 1-\frac{1}{\sqrt{2}}
\end{array},\ 
\begin{array}{c|c c c}
  0 & 0 & & \\
  2-\sqrt{2} & 1 - \frac{1}{\sqrt{2}} & 1 - \frac{1}{\sqrt{2}} & \\
  1 & \frac{1}{2\sqrt{2}} & \frac{1}{2\sqrt{2}} & 1 - \frac{1}{\sqrt{2}} \\
\hline
  & \frac{1}{2\sqrt{2}} & \frac{1}{2\sqrt{2}} & 1 - \frac{1}{\sqrt{2}}
\end{array}
\right),\label{eqn:ark2c}
\end{align}
with $a_{3,2} = \frac{1}{2}$, a third-order (four-stage) method (ARK 3), and a fourth-order (six-stage) method (ARK 4)
constructed in~\cite{kennedycarpenter}.
The implicit parts of the ARK methods used
here are ESDIRK (explicit first-stage, single-diagonal coefficient) and L-stable. The performance of the ARK methods is compared with that of the explicit RK methods: second-order, two-stage RK 2a, third-order, three-stage RK 3, and the classical fourth-order four-stage RK 4. Table~\ref{tab:methods} summarizes the time integration methods used in this paper.

\subsection{Linearization}
\label{subsec:linear}
The stage calculations \eqref{eqn:stages} require the solution of a nonlinear system of equations. We modify the partitioning of the RHS such that only a linear system needs to be solved instead. The fast term is linearized, and the implicit integrator is applied on this linear part. The slow term, redefined as total RHS with the linearized fast term subtracted from it, is treated explicitly. The linearized fast term removes the stiffness from the original RHS and reduces the computational cost of solving the implicit part. We note that the linearization does not introduce an error in the overall discretized equations. 

Ignoring the source term for now, we rewrite (\ref{eqn:stages}) as the following nonlinear system of equations for an implicit ARK stage:
\begin{align}
{\bf Q}^{\left(i\right)} - \sigma\hat{\bf F}_F\left({\bf Q}^{\left(i\right)}\right) &= {\bf Q}^n + \Delta t \sum_{j=1}^{i-1} \left\{ a_{ij} \hat{\bf F}_S\left({\bf Q}^{\left(j\right)}\right) + \tilde{a}_{ij} \hat{\bf F}_F\left({\bf Q}^{\left(j\right)}\right) \right\},\nonumber \\
\Rightarrow \left[\mathcal{I} - \sigma\mathcal{D}\otimes\mathcal{A}_F\right]{\bf Q}^{\left(i\right)} &= {\bf Q}^n + \Delta t \sum_{j=1}^{i-1} \left\{ a_{ij} \hat{\bf F}_S\left({\bf Q}^{\left(j\right)}\right) + \tilde{a}_{ij} \hat{\bf F}_F\left({\bf Q}^{\left(j\right)}\right) \right\},\label{eqn:imp_stage}
\end{align}
where $\sigma = \Delta t \tilde{a}_{ii}$. The nonlinearity of (\ref{eqn:imp_stage}) arises from two sources: the fast Jacobian $\mathcal{A}_F = \mathcal{A}_F\left({\bf Q}\right)$ and the WENO5/CRWENO5 finite-difference operator $\mathcal{D} = \mathcal{D}\left(\omega\right)$, where $\omega = \omega\left({\bf f}\left({\bf q}\right)\right)$ are the solution-dependent weights given by (\ref{eqn:weno_weights}).

The fast Jacobian is evaluated at the beginning of the step and kept
fixed for all the stages. The partitioning of the flux at stage $i$ is modified as follows:
\begin{align}
{\bf f}_F\left({\bf Q}^{\left(i\right)}\right) = \mathcal{A}_F\left({\bf Q}^n\right) {\bf Q}^{\left(i\right)},\ \ \ \ {\bf f}_S\left({\bf Q}^{\left(i\right)}\right) = {\bf f}\left({\bf Q}^{\left(i\right)}\right) - {\bf f}_F\left({\bf Q}^{\left(i\right)}\right).\label{eqn:linear_part1}
\end{align}
The corresponding expressions for spatially discretized partitioned flux terms are
\begin{align}
\hat{\bf F}_F\left({\bf Q}^{\left(i\right)}\right) &= \left[\mathcal{D} \otimes \mathcal{A}_F\left({\bf Q}^n\right) \right] {\bf Q}^{\left(i\right)},\nonumber\\
\hat{\bf F}_S\left({\bf Q}^{\left(i\right)}\right) &= 
\hat{\bf F}\left({\bf Q}^{\left(i\right)}\right) - \hat{\bf F}_F\left({\bf Q}^{\left(i\right)}\right)
= \left[\mathcal{D} \otimes \left\{\mathcal{A} \left({\bf Q}^{\left(i\right)}\right) - \mathcal{A}_F\left({\bf Q}^{n}\right)\right\}\right] {\bf Q}^{\left(i\right)}\label{eqn:linear_part2}.
\end{align}
Equation (\ref{eqn:linear_part2}) satisfies $\hat{\bf F}_F\left({\bf Q}^{\left(i\right)}\right) + \hat{\bf F}_S\left({\bf Q}^{\left(i\right)}\right) = \hat{\bf F}\left({\bf Q}^{\left(i\right)}\right)$ exactly. Therefore, the linearized partitioning is consistent with the unpartitioned RHS and does not introduce an error in the overall algorithm.

The nonlinear finite-difference operator $\mathcal{D}\left(\omega\right)$ is linearized by computing and fixing the solution-dependent weights (\ref{eqn:weno_weights}) at the beginning of each stage. The computation of ${\bf F}\left({\bf Q}^{\left(i\right)}\right)$ and ${\bf F}_F\left({\bf Q}^{\left(i\right)}\right)$ during the iterative solution of (\ref{eqn:imp_stage}) does not recalculate the weights $\omega$ based on the smoothness of the current guess for ${\bf Q}^{\left(i\right)}$. We define the finite-difference operator at stage $i$ as
\begin{equation}
\mathcal{\bar{D}} = \mathcal{D}\left(\bar{\omega}\right),\ {\rm where}\ \bar{\omega} = \left\{\begin{array}{lc} \omega\left({\bf f}\left({\bf Q}^{\left(i-1\right)}\right)\right) & i > 1 \\ \omega\left({\bf f}\left({\bf Q}^n\right)\right) & i = 1\end{array}\right.\label{eqn:linear_weights}.
\end{equation}
Thus, during the stage computation, the interpolation coefficients in (\ref{eqn:weno5}) or (\ref{eqn:crweno5}) are constant, and the resulting operators are linear.

Inspection of (\ref{eqn:hyp_cons}) shows that the source term is linear if the gravitational forces do not depend on the solution (this is true for our application). As previously mentioned, a well-balanced formulation~\cite{ghoshconstaAIAAJ2016} is used to evaluate it on the discretized domain; this formulation preserves its linearity. Denoting $\mathcal{S} = \partial{\bf S}/\partial{\bf Q}$ as the Jacobian of the source term, we apply (\ref{eqn:linear_part2}) and (\ref{eqn:linear_weights}) to (\ref{eqn:stages}) to obtain the following linear system of equations for the implicit ARK stages:
\begin{align}
\left[ \mathcal{I}  - \sigma\left\{\mathcal{\bar{D}}\otimes\mathcal{A}_F\left({\bf Q}^n\right)+\mathcal{S}\right\}\right]&{\bf Q}^{\left(i\right)} \nonumber\\
= {\bf Q}^n + \Delta t \sum_{j=1}^{i-1} &\left\{ a_{ij} \hat{\bf F}_S\left({\bf Q}^{\left(j\right)}\right) + \tilde{a}_{ij} \left[\mathcal{\bar{D}}\otimes\mathcal{A}_F\left({\bf Q}^n\right)+\mathcal{S}\right]{\bf Q}^{\left(j\right)} \right\},\label{eqn:imp_stage_linear}
\end{align}
where $\hat{\bf F}_S$ is defined by (\ref{eqn:linear_part2}). Equation (\ref{eqn:imp_stage_linear}) is solved iteratively by using the generalized residual method (GMRES)~\cite{saad,gmres} implemented in the Krylov solver module (KSP) of PETSc, and a Jacobian-free approach is adopted where the Jacobian
\begin{equation}
\mathcal{J} \equiv \left[ \mathcal{I}  - \sigma\mathcal{\bar{D}}\otimes\mathcal{A}_F\left({\bf Q}^n\right) + \mathcal{S}\right]\label{eqn:jac_mf}
\end{equation}
is specified as its action on a vector. The stopping criterion for the iterative solver is specified as
$\| {\bf r}_{k+1} - {\bf r}_{k} \|_2 \le \max \left( \tau_r \|{\bf r}_0 \|_2 ,\tau_a \right)$,
where $\tau_a$ and $\tau_r$ are the absolute and relative tolerances, respectively; ${\bf r}$ is the residual given by
\begin{align}
{\bf r}_k = &\left[ \mathcal{I}  - \sigma\left\{\mathcal{\bar{D}}\otimes\mathcal{A}_F\left({\bf Q}^n\right)+\mathcal{S}\right\}\right]{\bf Q}^{\left(i\right)}_k \nonumber\\
&- \left[ {\bf Q}^n + \Delta t \sum_{j=1}^{i-1} \left\{ a_{ij} \hat{\bf F}_S\left({\bf Q}^{\left(j\right)}\right) + \tilde{a}_{ij} \left[\mathcal{\bar{D}}\otimes\mathcal{A}_F\left({\bf Q}^n\right)+\mathcal{S}\right]{\bf Q}^{\left(j\right)} \right\} \right];\label{eqn:residual}
\end{align}
and the subscript $k$ denotes the $k$th guess for the stage solution ${\bf Q}^{\left(i\right)}$.

\subsection{Modified Upwinding}
\label{subsec:consistency}

The interpolated flux at a grid interface is computed by using (\ref{eqn:rusanov}), which can be written for the total and the fast flux terms as follows:
\begin{align}
\hat{\bf f}_{j+1/2} &= \frac{1}{2} \left[ \hat{\bf f}^L_{j+1/2} + \hat{\bf f}^R_{j+1/2} - \delta_{j+1/2} \left( \hat{\bf q}^R_{j+1/2} - \hat{\bf q}^L_{j+1/2} \right) \right],\label{eqn:upwind_flux} \\
\hat{\bf f}_{F,j+1/2} &= \frac{1}{2} \left[ \hat{\bf f}^L_{F,j+1/2} + \hat{\bf f}^R_{F,j+1/2} - \delta^F_{j+1/2} \left( \hat{\bf q}^R_{j+1/2} - \hat{\bf q}^L_{j+1/2} \right) \right]\label{eqn:upwind_fast_flux},
\end{align}
where $\delta,\delta^F$ are the diffusion coefficients for the upwinding scheme and $\hat{\bf f}, \hat{\bf f}_F$ are the reconstructed numerical total and fast flux terms at the grid interfaces, related to $\hat{\bf F}, \hat{\bf F}_F$ in (\ref{eqn:linear_part2}) through (\ref{eqn:cons_fd}). Subtracting (\ref{eqn:upwind_fast_flux}) from (\ref{eqn:upwind_flux}) results in the following expression for the slow flux at a grid interface:
\begin{align}
\hat{\bf f}_{S,j+1/2} =& \frac{1}{2} \left[ \left( \hat{\bf f}^L_{j+1/2} + \hat{\bf f}^R_{j+1/2} \right) - \left( \hat{\bf f}^L_{F,j+1/2} + \hat{\bf f}^R_{F,j+1/2} \right) \right] \nonumber\\ 
&- \frac{1}{2}\left[\left( \delta_{j+1/2} - \delta^F_{j+1/2} \right) \left( \hat{\bf q}^R_{j+1/2} - \hat{\bf q}^L_{j+1/2} \right) \right].\label{eqn:upwind_slow_flux}
\end{align}
If the same diffusion coefficient for the upwinding scheme is used for both total and the fast flux,
\begin{equation}
\delta_{j+1/2} = \delta^F_{j+1/2} =  \max_{j,j+1} \nu,
\end{equation}
we obtain a central discretization of the slow flux term (\ref{eqn:upwind_slow_flux}) with no diffusion and purely imaginary eigenvalues. This is undesirable with respect to the ARK time integrator, as explained in Sec.~\ref{subsec:lin:stab}. To avoid this, we modify the upwinding method to apply the diffusion specifically along the characteristic fields that the flux term represents. The diffusion coefficients are expressed as
\begin{equation}
\left[\tilde{\delta}\right]_{j+1/2} = \mathcal{X}\left[\begin{array}{ccc} \bar{\mu} & & \\ & \bar{\nu} & \\ & & \bar{\nu} \end{array}\right]\mathcal{X}^{-1},\ \ \ \ \left[\tilde{\delta}^F\right]_{j+1/2} = \mathcal{X}\left[\begin{array}{ccc} 0 & & \\ & \bar{\nu} & \\ & & \bar{\nu} \end{array}\right]\mathcal{X}^{-1},
\end{equation}
where
\begin{equation}
\bar{\nu} = \max_{j,j+1} \left( \left|u\right| + a \right),\ \ \ \ \bar{\mu} = \max_{j,j+1} \left|u\right|,
\end{equation}
and the equations to compute the flux at the grid interfaces from their left- and right-biased interpolated values are
\begin{align}
\hat{\bf f}_{j+1/2} &= \frac{1}{2} \left\{ \hat{\bf f}^L_{j+1/2} + \hat{\bf f}^R_{j+1/2} - \left[\tilde{\delta}\right]_{j+1/2} \left( \hat{\bf q}^R_{j+1/2} - \hat{\bf q}^L_{j+1/2} \right) \right\},\label{eqn:upwind_flux_mod} \\
\hat{\bf f}_{F,j+1/2} &= \frac{1}{2} \left\{ \hat{\bf f}^L_{F,j+1/2} + \hat{\bf f}^R_{F,j+1/2} - \left[\tilde{\delta}^F\right]_{j+1/2} \left( \hat{\bf q}^R_{j+1/2} - \hat{\bf q}^L_{j+1/2} \right) \right\} \label{eqn:upwind_fast_flux_mod}.
\end{align}
We then obtain the following expression for the slow term:
\begin{equation}
\hat{\bf f}_{S,j+1/2} = \frac{1}{2} \left\{ \hat{\bf f}^L_{F,j+1/2} + \hat{\bf f}^R_{F,j+1/2} - \left[\tilde{\delta}^S\right]_{j+1/2} \left( \hat{\bf q}^R_{j+1/2} - \hat{\bf q}^L_{j+1/2} \right) \right\} \label{eqn:upwind_slow_flux_mod},
\end{equation}
where
\begin{equation}
\left[\tilde{\delta}^S\right]_{j+1/2} = \left[\tilde{\delta}\right]_{j+1/2} - \left[\tilde{\delta}^F\right]_{j+1/2} = \mathcal{X}\left[\begin{array}{ccc} \bar{\mu} & & \\ & 0 & \\ & & 0 \end{array}\right]\mathcal{X}^{-1}.
\end{equation}
The modified upwinding applies the diffusion to the fast term only along the acoustic modes and to the slow term only along the advective mode; it does not add any additional diffusion compared with the spatial discretization of the unsplit flux. This modified upwinding scheme resembles the Roe upwinding scheme~\cite{roe}.

\subsection{Linear Stability Considerations}
\label{subsec:lin:stab}
We analyze the linear stability of the semi-implicit time integration method
(\ref{eq:ARK}) by considering a linear test problem,
\begin{align}
\label{eq:ODE:linear}
&
\mathbf{Q}'(t) = \lambda \mathbf{Q}(t) + \mu \mathbf{Q}(t)\,,
\end{align}
where $\lambda,\,\mu \in\mathbb{C}$ represent eigenvalues of the
nonstiff (slow) and stiff (fast) components \cite{constasandu2010},
respectively, and $\mathbb{C}$ is the set of complex numbers.
In our application, $\lambda\mathbf{Q}(t)$ represents the 
slow component $\hat{\bf F}_S\left({\bf Q}\right)$, and 
$\mu\mathbf{Q}(t)$ represents the fast component 
$\hat{\bf F}_F\left({\bf Q}\right)+\hat{\bf S}\left({\bf Q}\right)$.
This problem provides useful insights into the stability
behavior for the nonlinear problem. A time step can be expressed as
\begin{align}
\label{eq:ODE:timestep}
&
\mathbf{Q}^{n+1}= R(\lambda \dt,\mu \dt)\,\mathbf{Q}^{n}\,, 
\end{align}
where $R$ is the stability function of the method. The stability region ${\cal S}$ of the
semi-implicit method is then defined by 
\begin{align}
\label{eq:ODE:stabreg}
&
{\cal S}=\left\{\lambda  \dt \in
\mathbb{C},\,\mu \dt  \in  \mathbb{C}: | R(\lambda \dt,\mu \dt) 
| \le 1 \right\} \subset~ \mathbb{C} \times \mathbb{C}.
\end{align}
The high dimensionality of the stability region makes its analysis 
difficult.
To simplify it, we fix the stiff stability region ${\cal S}_\mu$ as
the set of stiff eigenvalues 
\begin{align}
&
{\cal S}_\mu = \{\mu_1 \dt,\mu_2 \dt, \dots,\mu_k\dt\}\,,
\end{align}
where $k$ is the total number of eigenvalues, and focus on the nonstiff 
stability region ${\cal S}_\lambda \subset \mathbb{C}$. 
The method is stable for all $\lambda \dt \in {\cal S}_\lambda$ 
for the nonstiff component subject to ${\cal S}_\mu$. 

The condition $\Re\left(\lambda\right)\rightarrow 0$ imposes
tight restrictions on the classes of methods that can be used in
practice because of the stability properties of the time integration
methods. 
It is challenging to construct methods whose ${\cal S}_\lambda$ has
a large overlap with the imaginary axis in the complex plane. 
Moreover, the overlap of ${\cal S}_\lambda$ with the imaginary axis
is negatively impacted for $\Re\left(\mu\right)\rightarrow 0$~\cite{constasandu2010}, 
as is the case in our application. Semi-implicit methods with explicit imaginary stability
that are less dependent on the implicit operator have been
constructed~\cite{kennedycarpenter,durranblossey,giraldokellyconsta2013};
however, relaxing this restriction allows for more efficient
methods. 

\begin{figure}[t!]
\begin{center}
\subfigure[$a_{3,2}=\frac{1}{6}(3+2\sqrt{2})$ ]{\includegraphics[width=0.49\textwidth]{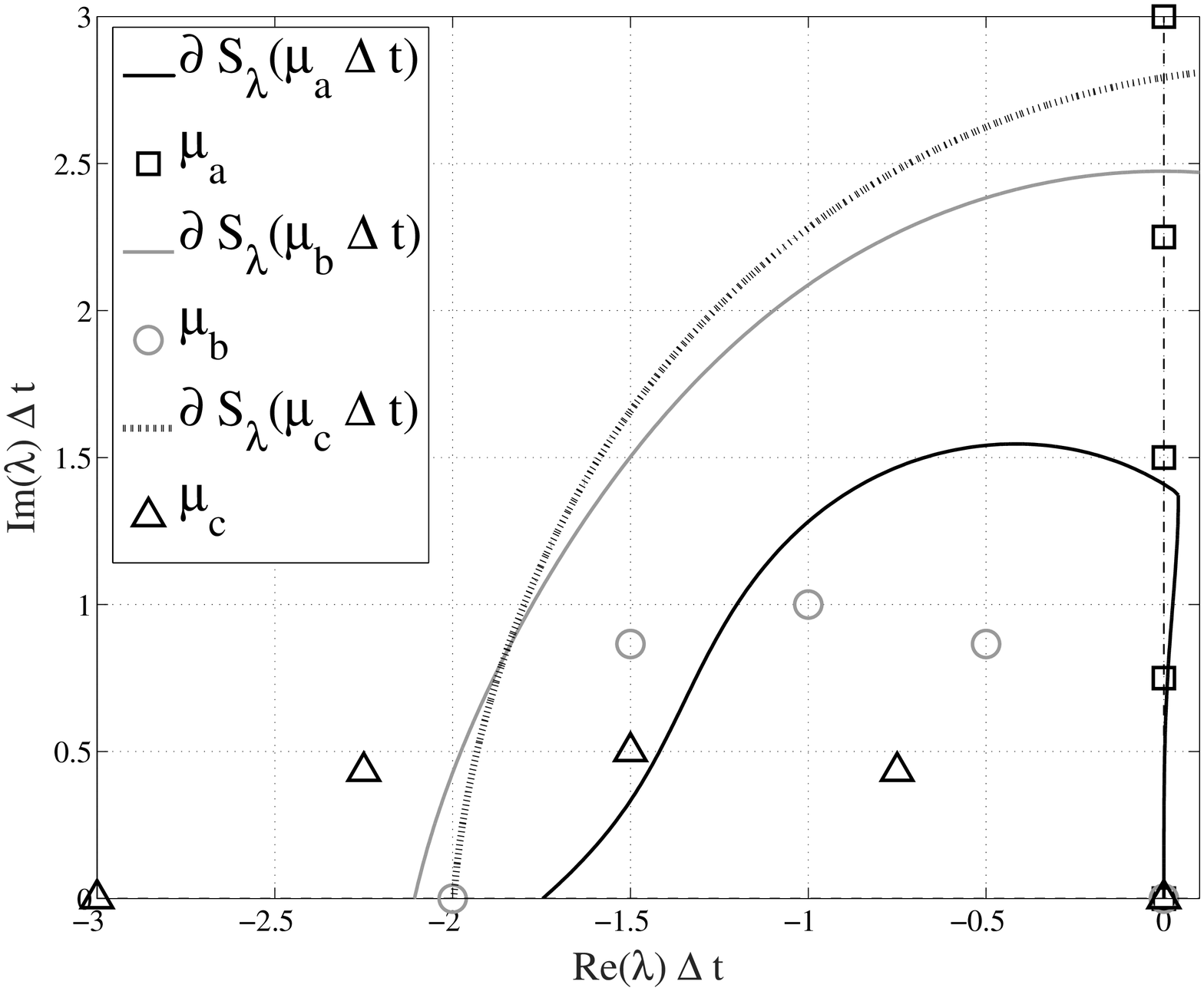}\label{fig:stab:2e}}
\subfigure[$a_{3,2}=\frac{1}{2}$ ]{\includegraphics[width=0.49\textwidth]{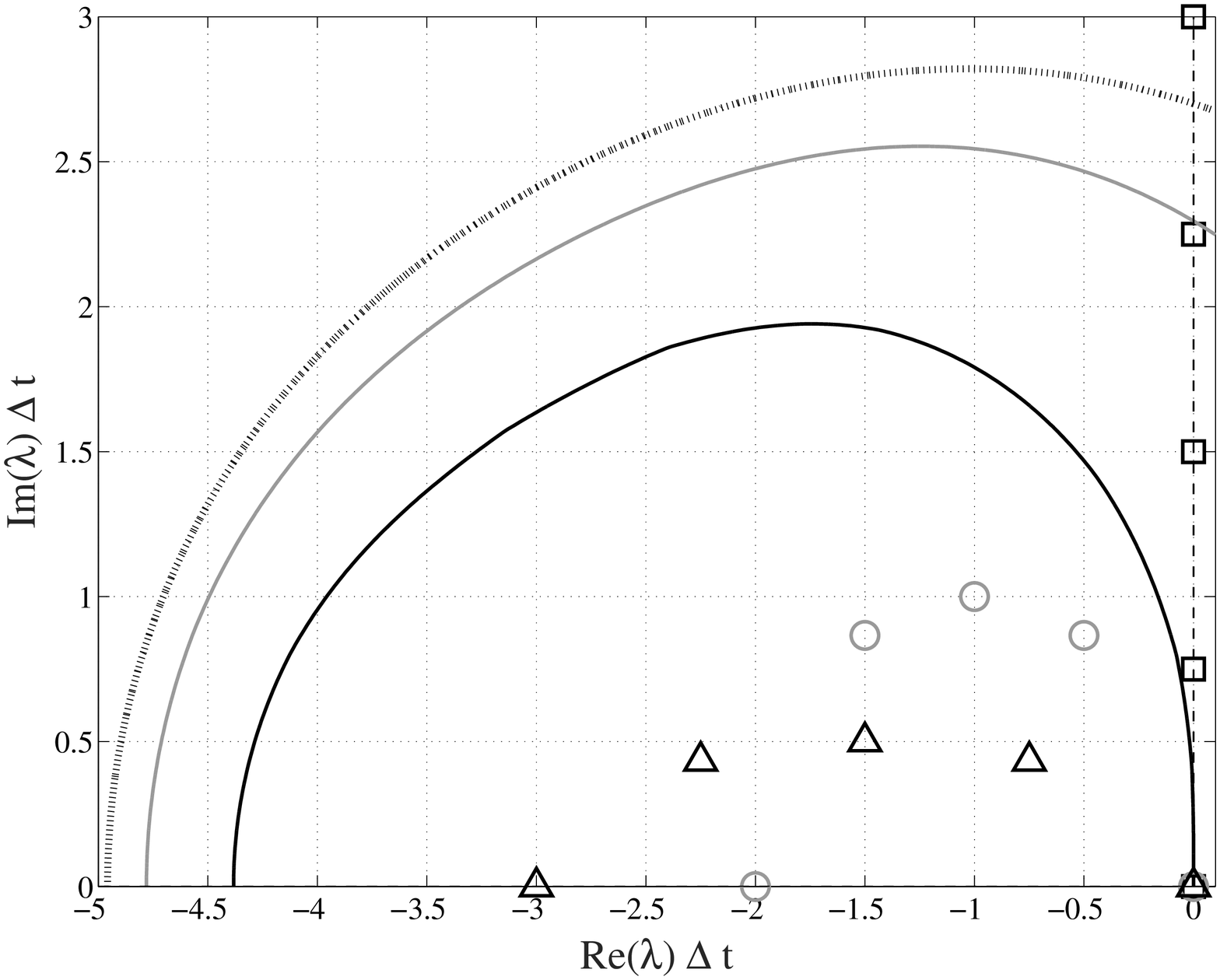}\label{fig:stab:2c}}
\end{center}
\caption{The (explicit) stability regions for method \eqref{eqn:ark2c} for
  three fixed stiff eigenvalues sets with different values for
  $a_{3,2}$ coefficients. The stability region is degraded more with
  method coefficients set in \ref{fig:stab:2e} than with
  \ref{fig:stab:2c} as the implicit eigenvalues are set to be pure imaginary.}
\label{fig:stab}
\end{figure}

Figure~\ref{fig:stab} illustrates the behavior of the stability region
for scheme \eqref{eqn:ark2c} using $a_{3,2}=\frac{1}{6}(3+2\sqrt{2})$
and $a_{3,2}=\frac{1}{2}$ for different sets of implicit
eigenvalues. The stiff eigenvalues $\mu_{\left(\cdot\right)}$
and the boundaries of the corresponding explicit stability regions 
$\partial {\cal S}_\lambda\left(\mu_{\left(\cdot\right)}\Delta t\right)$
are shown; the subscript $a$ refers to the case where $\mu$ are purely imaginary,
the subscript $b$ refers to the case where $\mu$ has both real and imaginary
components, and the subscript $c$ refers to the case where the real part 
of $\mu$ is larger than its imaginary part. 
Overall, the size of the stability region
reduces as the imaginary components of the stiff eigenvalues increase.
In addition, the overlap of
${\cal S}_\lambda\left(\mu_{\left(\cdot\right)}\Delta t\right)$ with the 
imaginary axis is largest
for $\mu_c$ and smallest for $\mu_a$. 
The degradation of ${\cal S}_\lambda$ is more pronounced in Fig.~\ref{fig:stab:2e},
and thus we choose $a_{3,1}=a_{3,2}=\frac{1}{2}$ in \eqref{eqn:ark2c}. A more detailed
discussion is presented in \cite{giraldokellyconsta2013}. This brief analysis demonstrates the
importance of avoiding imaginary eigenvalues $\lambda$ for the nonstiff component $\hat{\bf F}_S\left({\bf Q}\right)$
since $\Re\left(\mu\right)\rightarrow0$ holds true for several eigenvalues of $\hat{\bf F}_F\left({\bf Q}\right)+\hat{\bf S}\left({\bf Q}\right)$.

\subsection{Preconditioning}
\label{subsec:precon}

The block Jacobi preconditioner~\cite{saad}, implemented in the preconditioning module (PC) of PETSc, is used in the current work. Although the Jacobian of the implicitly treated operator is specified in a matrix-free way (\ref{eqn:jac_mf}), an approximation to the Jacobian is provided as a sparse matrix. The approximate Jacobian for the preconditioner is defined as
\begin{equation}
\mathcal{J}_p \equiv \left[ \mathcal{I}  - \sigma\mathcal{\bar{D}_{\rm 1st}}\otimes\mathcal{A}_F\left({\bf Q}^n\right) + \mathcal{S}\right] \approx \mathcal{J}\label{eqn:jac_approx},
\end{equation}
where $\mathcal{D}_{\rm 1st}$ represents a first-order upwind discretization operator. This results in a block tridiagonal matrix for the one-dimensional system and will result in block penta- and septa-diagonal systems for two- and three-dimensional flows, respectively. Development of more advanced preconditioning techniques for the algorithm presented here is beyond the scope of this paper and will be studied in the future.

\section{Extension to Two-Dimensional Flows}
\label{sec:2d}

The two-dimension Euler equations with gravitational source terms can be expressed as the following hyperbolic conservation law:
\begin{equation}\label{eqn:hyp_cons_2d}
\frac{\partial {\bf q}}{\partial t} + \frac{\partial {\bf f}\left({\bf q}\right)}{\partial x} + \frac{\partial {\bf h}\left({\bf q}\right)}{\partial y} = {\bf s}\left({\bf q}\right),
\end{equation}
where
\begin{equation}
{\bf q} = \left[\begin{array}{c} \rho \\ \rho u \\ \rho v \\ e \end{array}\right],
{\bf f} = \left[\begin{array}{c} \rho u \\ \rho u^2 + p \\ \rho u v \\ (e+p) u \end{array}\right],
{\bf h} = \left[\begin{array}{c} \rho v \\ \rho u v \\ \rho v^2 + p \\ (e+p) v \end{array}\right],
{\bf s} = \left[\begin{array}{c} 0 \\ - \rho {\bf g}\cdot\hat{\bf i} \\  - \rho {\bf g}\cdot\hat{\bf j}  \\ - \left(\rho u  {\bf g}\cdot\hat{\bf i} + \rho v {\bf g}\cdot\hat{\bf j} \right)  \end{array}\right].\nonumber
\end{equation}
The Cartesian unit vectors along $x$ and $y$ are denoted by $\hat{\bf i}$ and $\hat{\bf j}$, respectively, and $u,v$ are the velocity components along $x,y$. The spatial discretization described in the preceding sections is extended to the two-dimensional equations through a dimension-by-dimension approach, where the derivatives along one dimension are computed independently of the other dimension. This paper considers only problems solved on Cartesian grids. The eigenvalues of the two-dimensional system are given by
\begin{equation}
\Lambda\left[ \frac{\partial \left({\bf f},{\bf h}\right)}{\partial {\bf q}} \right] = \left\{\left(u,v\right), \left(u,v\right), \left(u,v\right)-a,\left(u,v\right)+a \right\},
\end{equation}
and they are split into their advective and acoustic components as
\begin{equation}
\Lambda_S = \left\{\left(u,v\right), \left(u,v\right), 0, 0 \right\},
\Lambda_F = \left\{0, 0, \left(u,v\right)-a,\left(u,v\right)+a \right\}.
\end{equation}
The slow and fast Jacobians are obtained by using the similarity transformations given by (\ref{eqn:fastslowjac}), and the partitioned flux and its spatially discretized counterpart are then computed. The left and right eigenvectors for the two-dimensional Euler equations are provided in~\cite{hirsch,normal-eigen}, and we use these in this paper. The resulting semi-discrete ODE can be expressed as
\begin{equation}\label{eqn:semidisc_ode_part_2d}
\frac{d {\bf Q}}{dt} = \left\{ \hat{\bf F}_F\left({\bf Q}\right) + \hat{\bf H}_F\left({\bf Q}\right) \right\} + \left\{ \hat{\bf F}_S\left({\bf Q}\right) + \hat{\bf H}_S\left({\bf Q}\right) \right\} + \hat{\bf S}\left({\bf Q}\right),
\end{equation}
where $\hat{\bf H}_{F,S}$ denotes the spatially discretized partitioned fluxes along the $y$-direction. Equation (\ref{eqn:semidisc_ode_part_2d}) is integrated in time by using an ARK method given by (\ref{eqn:ark}), where the fast flux terms and the source term $\left\{ \hat{\bf F}_F + \hat{\bf H}_F\right\} + \hat{\bf S}$ are treated implicitly and the slow flux terms $ \left\{ \hat{\bf F}_S + \hat{\bf H}_S\right\}$ are treated explicitly.

\section{Numerical Tests}
\label{sec:numtests}

The performance of the semi-implicit time integrators with characteristic-based flux partitioning is tested in this section with two simple flow problems. The tests verify that the integration of the acoustic modes in time by using an implicit method results in a largest stable time step that is determined by the advective scale. In addition, the accuracy and convergence of the time integration methods are demonstrated. The two problems solved in this section are formulated in terms of nondimensional flow variables.

\subsection{Density Wave Advection}
\label{subsec:densitywave1d}

\begin{figure}[t!]
\begin{center}
\subfigure[$M_\infty=0.1$ ]{\includegraphics[width=0.49\textwidth]{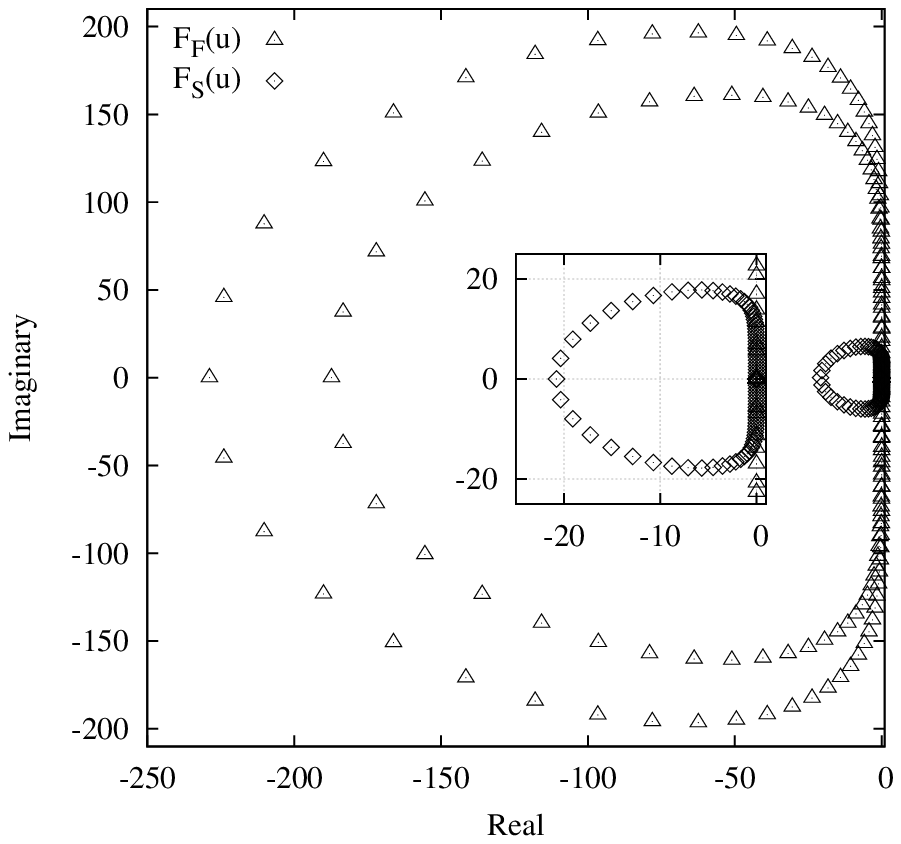}\label{fig:eigenvalues_1d_m0_10}}
\subfigure[$M_\infty=0.01$]{\includegraphics[width=0.49\textwidth]{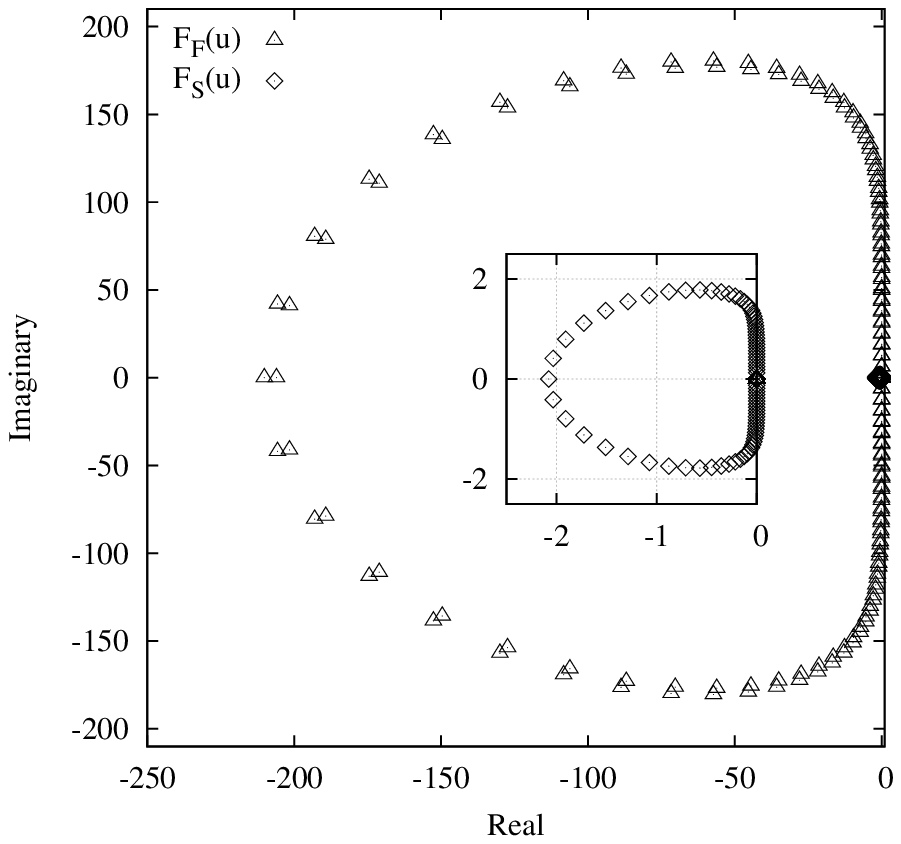}\label{fig:eigenvalues_1d_m0_01}}
\end{center}
\caption{Eigenvalues of Jacobians of the partitioned flux terms $\hat{\bf F}_{F,S}$ for the one-dimensional density wave advection at two Mach numbers. The CRWENO5 scheme is used, and the problem is discretized on a grid with $80$ points. The insets are magnified plots of the eigenvalues of the slow flux term $\hat{\bf F}_S$.}
\label{fig:eigenvalues_1d}
\begin{center}
\subfigure[$M_\infty=0.1$ ]{\includegraphics[width=0.49\textwidth]{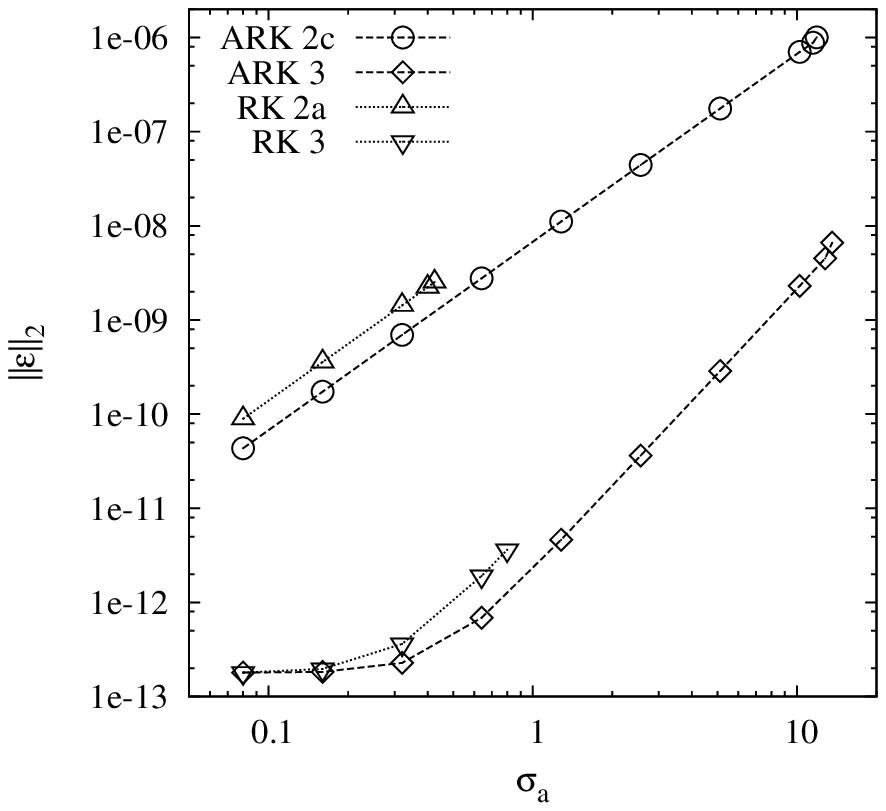}\label{fig:Euler1D_DensityWave_Err_Dt_0_10}}
\subfigure[$M_\infty=0.01$]{\includegraphics[width=0.49\textwidth]{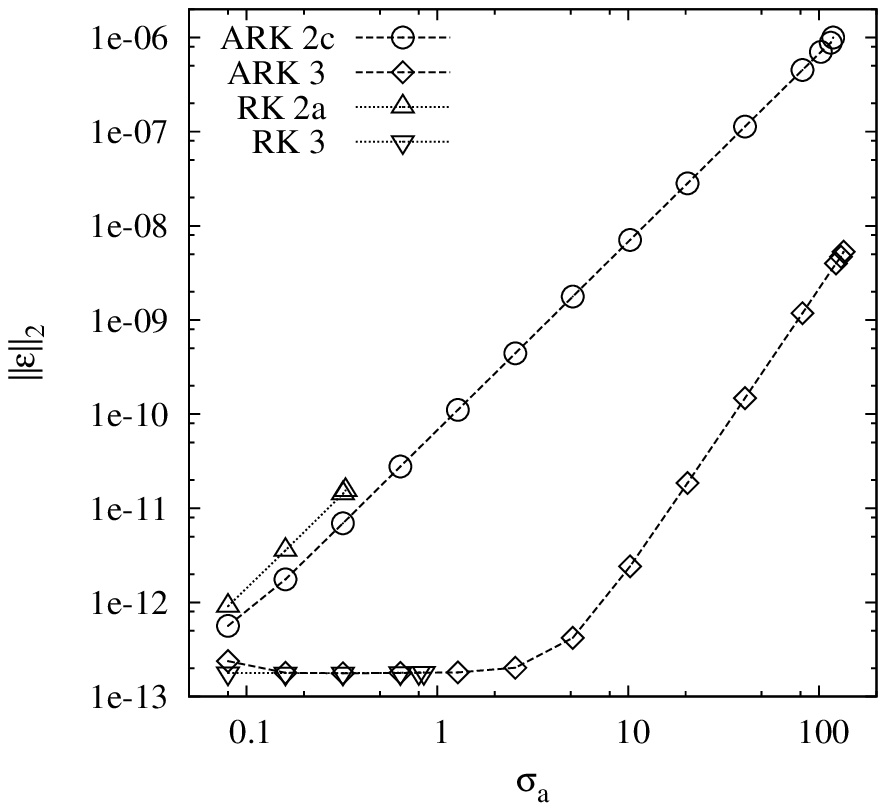}\label{fig:Euler1D_DensityWave_Err_Dt_0_01}}
\end{center}
\caption{Density wave advection: $L_2$ norm of the error as a function of the acoustic CFL number for the ARK and explicit RK methods.}
\label{fig:Euler1D_DensityWave_Err_Dt}
\end{figure}

This one-dimensional test problem involves the advection of a sinusoidal density wave over a periodic domain. The exact solution is given by
\begin{equation}\label{eqn:density_adv_soln}
\rho\left(x,t\right) = \rho_\infty + \hat{\rho}\sin\left[ 2 \pi \left(x - u_\infty t \right)\right],\ u\left(x,t\right) = u_\infty,\ p\left(x,t\right) = p_\infty.
\end{equation}
With this solution, the Euler equations are equivalent to the linear advection equation. The mean density and pressure are taken as $\rho_\infty = 1$ and $p_\infty = 1/\gamma$, respectively, resulting in the mean speed of sound as $a_\infty = 1$. We consider two values for the mean Mach number (given by $M_\infty = u_\infty/a_\infty$): $0.1$ and $0.01$. The domain is $x \in [0,1]$, and periodic boundary conditions are applied at the boundaries. 

Figure~\ref{fig:eigenvalues_1d} shows the eigenvalues of the partitioned flux Jacobians for the CRWENO5 scheme on a grid with $80$ points. The Jacobians are evaluated from the discretized operators $\hat{\bf F}_{F,S}$ through finite differences. The eigenvalues for the case with mean Mach number $0.1$ is shown in Figure~\ref{fig:eigenvalues_1d_m0_10}, with the magnified subplot showing the eigenvalues of the slow flux term $\hat{\bf F}_S$. As shown earlier, the flux partitioning results in a separation of the advective and acoustic modes. The eigenvalues of the slow flux correspond to the advective mode, and they are smaller in magnitude than those of the fast flux by an approximate factor of $10$ (the inverse of the Mach number). Figure~\ref{fig:eigenvalues_1d_m0_01} shows the eigenvalues for the case with a mean Mach number of $0.01$.  At this smaller Mach number, the separation between the advective and acoustic scales is larger. The magnitudes of the eigenvalues of the slow flux are smaller than those of the fast flux by an approximate factor of $M_\infty^{-1} = 100$. The two acoustic modes are characterized by the wave speeds $u\pm a$; and thus, as the Mach number decreases, they converge to $a$.

Figure~\ref{fig:Euler1D_DensityWave_Err_Dt} shows the error as a function of the acoustic Courant-Friedrichs-Lewy (CFL) for the second- and third-order ARK methods, ARK 2c and ARK 3, as well as the two explicit Runge-Kutta (RK) methods of the same orders, RK 2a and RK 3. The solutions are obtained with the tolerances for the iterative solver specified as $\tau_r = \tau_a = 10^{-10}$. The final times for both cases correspond to one cycle over the periodic domain ($10$ for $M_\infty = 0.1$ and $100$ for $M_\infty = 0.01$). The time step sizes are increased from a small value until they reach a value for which the solution blows up, thus indicating the largest stable time step size of that time integrator. The error and the acoustic CFL are defined as
\begin{equation}
{\bf \epsilon} = {\bf Q}\left(x,t\right) - {\bf Q}_{\rm exact}\left(x,t\right),\ \ \ \ \sigma_a = a_\infty \frac{\Delta t}{\Delta x},
\end{equation}
where ${\bf Q}_{\rm exact}$ is given by (\ref{eqn:density_adv_soln}). The ARK methods converge at their theoretical orders for all the cases. The case with $M_\infty = 0.1$ is shown in Figure~\ref{fig:Euler1D_DensityWave_Err_Dt_0_10}, and the largest stable time steps for the ARK methods are observed to be larger than those of the explicit RK methods by a factor of approximately $M_\infty^{-1} = 10$. The explicit RK methods are restricted in their time step size by the acoustic mode, while the implicit treatment of the acoustic modes in the ARK method allows time step sizes restricted by the advective mode. Figure~\ref{fig:Euler1D_DensityWave_Err_Dt_0_01} shows the case with $M_\infty = 0.01$. The advective eigenvalues are smaller in magnitude for this lower Mach number, and therefore larger time step sizes are possible. The largest time step sizes for the ARK methods are again approximately $M_\infty^{-1} = 100$ times larger than those of the explicit RK methods. This demonstrates that the stability limits for the ARK methods are determined by the advective time scale because of the characteristic-based flux partitioning.

\subsection{Isentropic Vortex Convection}
\label{subsec:isen_vort}

\begin{figure}[t!]
\begin{center}
\subfigure[Eigenvalues]{\includegraphics[width=0.49\textwidth]{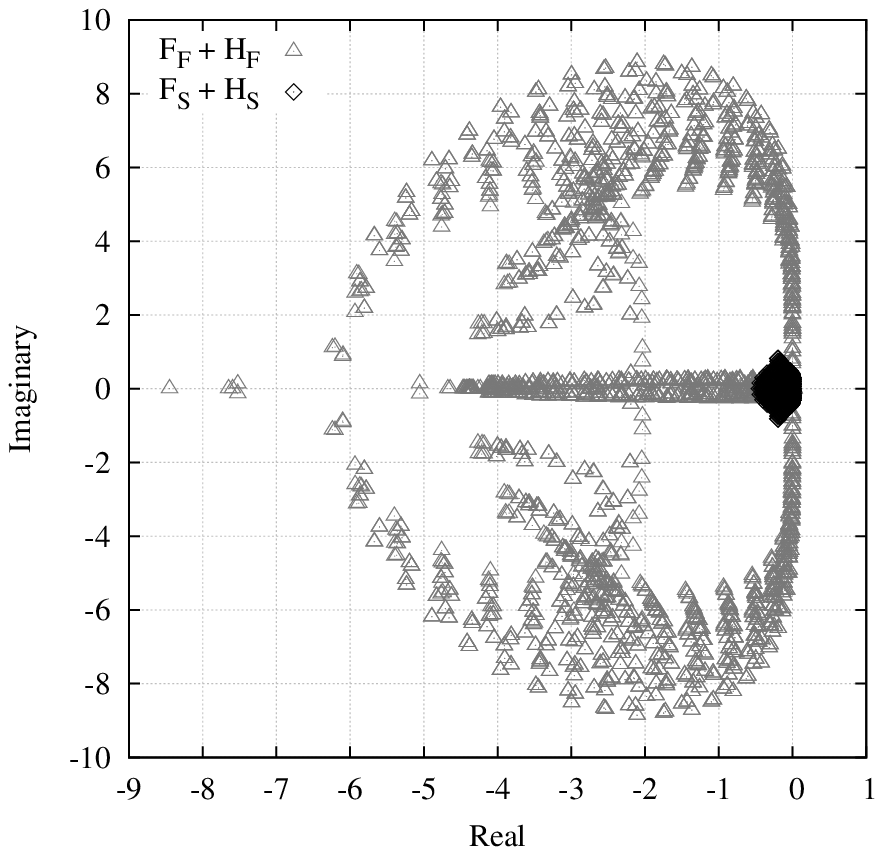}\label{fig:eigenvalues_vortconv}}
\subfigure[Magnified plot of the advective eigenvalues]{\includegraphics[width=0.49\textwidth]{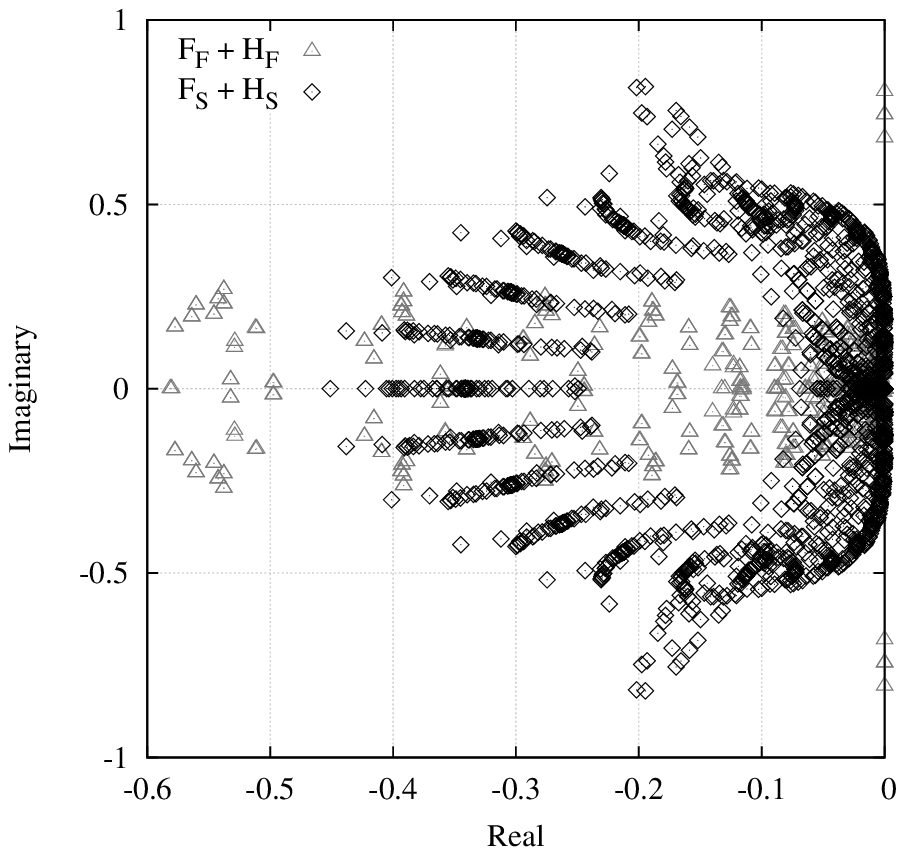}\label{fig:eigenvalues_vortconv_zoom}}
\end{center}
\caption{Eigenvalues of Jacobians of the partitioned flux terms $\left(\hat{\bf F}_F+\hat{\bf H}_F\right)$ and $\left(\hat{\bf F}_S+\hat{\bf H}_S\right)$ in (\ref{eqn:semidisc_ode_part_2d}) for the isentropic vortex convection case. The WENO5 scheme is used, and the problem is discretized on a grid with $32^2$ points.}
\label{fig:eigenvalues_2d_vortconv}
\end{figure}
\begin{figure}[h!]
\begin{center}
\subfigure[ARK 4, $\sigma_a\approx7.6$]{\includegraphics[width=0.49\textwidth]{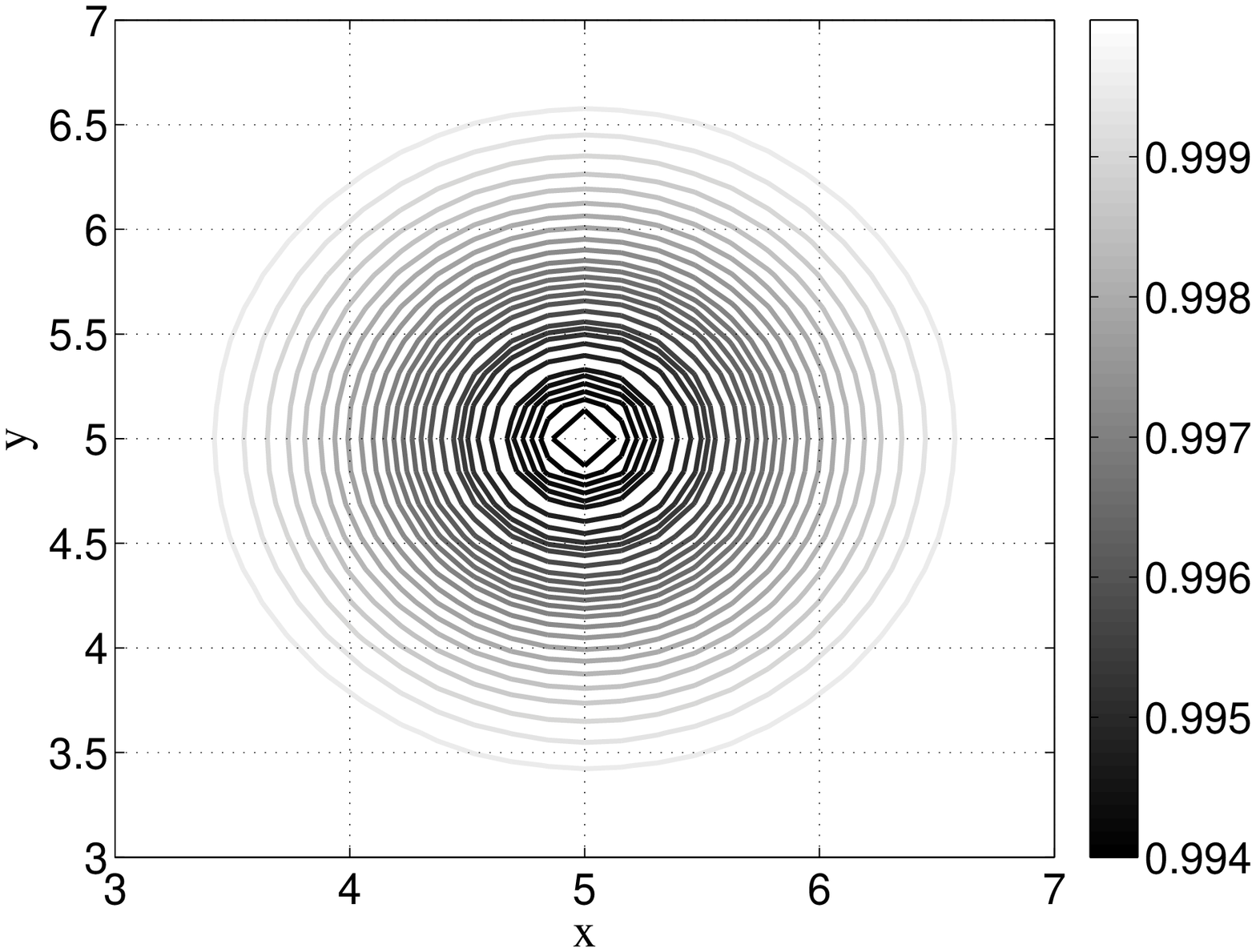}\label{fig:VortConv_Contour_dt01.000}}
\subfigure[Cross-sectional density ($y=5$)]{\includegraphics[width=0.49\textwidth]{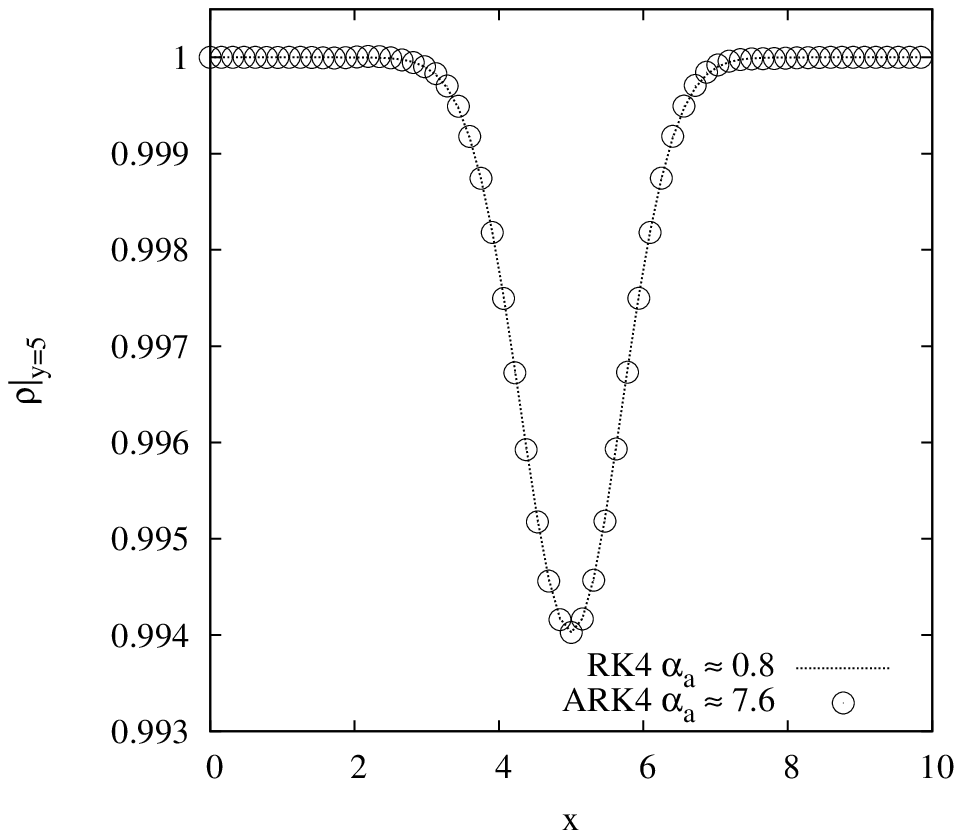}\label{fig:VortConv_CrossSc}}
\end{center}
\caption{Density contours of the isentropic vortex convection case after $2$ cycles over the periodic domain obtained with the WENO5 scheme on a grid with $64^2$ points, and the cross-sectional density profile at $y=5$; $\sigma_a$ is the acoustic CFL number.}
\label{fig:VortConv_Contour}
\end{figure}

The convection of an isentropic vortex~\cite{shuicase} is used to test the flux partitioning in two dimensions. The flow involves the inviscid convection of a vortex over a periodic domain and tests the ability of the numerical method to preserve the shape and strength of the vortex. We modify the original test case by reducing the Mach number. The domain is specified as $\left(x,y\right) \in \left[0,10\right]^2$, and the mean (freestream) flow is $\rho_\infty = 1$, $\ u_\infty = 0.1$, $\ v_\infty = 0,$ and $\ p_\infty = 1$.
A vortex is introduced in the flow, whose density and pressure are specified as
\begin{align}
\rho = \left[ 1 - \frac{\left(\gamma-1\right)b^2}{8\gamma\pi^2} e^{1-r^2} \right]^{\frac{1}{\gamma-1}},\ p = \rho^\gamma,
\end{align}
and thus $\rho,p \rightarrow \rho_\infty,p_\infty$ as $r \rightarrow \infty$. The velocity field is
\begin{align}
u = u_\infty - \frac{b}{2\pi} e^{\frac{1}{2}\left(1-r^2\right)} \left(y-y_c\right),\ v = v_\infty + \frac{b}{2\pi} e^{\frac{1}{2}\left(1-r^2\right)} \left(x-x_c\right),
\end{align}
where $b=0.5$ is the vortex strength and $r = \left[(x-x_c)^2 + (y-y_c)^2 \right]^{1/2}$ is the distance from the vortex center $\left(x_c,y_c\right) = \left(5,5\right)$. Periodic boundary conditions are applied at all boundaries. As the solution is evolved in time, the vortex convects over the periodic domain with a time period of $T_p = 100$. 

Figure~\ref{fig:eigenvalues_2d_vortconv} shows the eigenvalues of the partitioned Jacobians for the WENO5 scheme on a grid with $32^2$ points. The freestream Mach number for this example is $M_\infty \approx 0.08$, and thus we see a significant separation in the magnitudes of the advective and acoustic eigenvalues. Figure~\ref{fig:eigenvalues_vortconv_zoom} is a magnified plot of the advective eigenvalues. This demonstrates that the extension of the characteristic-based partitioning to two dimensions, as described in Section~\ref{sec:2d}, works as expected. Figure~\ref{fig:VortConv_Contour_dt01.000} shows the density contours of the flow for the solution obtained with the ARK 4 method at an acoustic CFL number of $\sim7.6$ on a grid with $64^2$ points and the WENO5 scheme. The final time is $200$, corresponding to $2$ cycles over the periodic domain. The horizontal cross-sectional density profile through $y=5$ for these solutions is shown in Figure~\ref{fig:VortConv_CrossSc}. The solution obtained with ARK 4 agrees well with that obtained with the explicit RK 4 scheme at an acoustic CFL number of $0.8$.

\begin{figure}[t!]
\begin{center}
\subfigure[Error vs. acoustic CFL]{\includegraphics[width=0.49\textwidth]{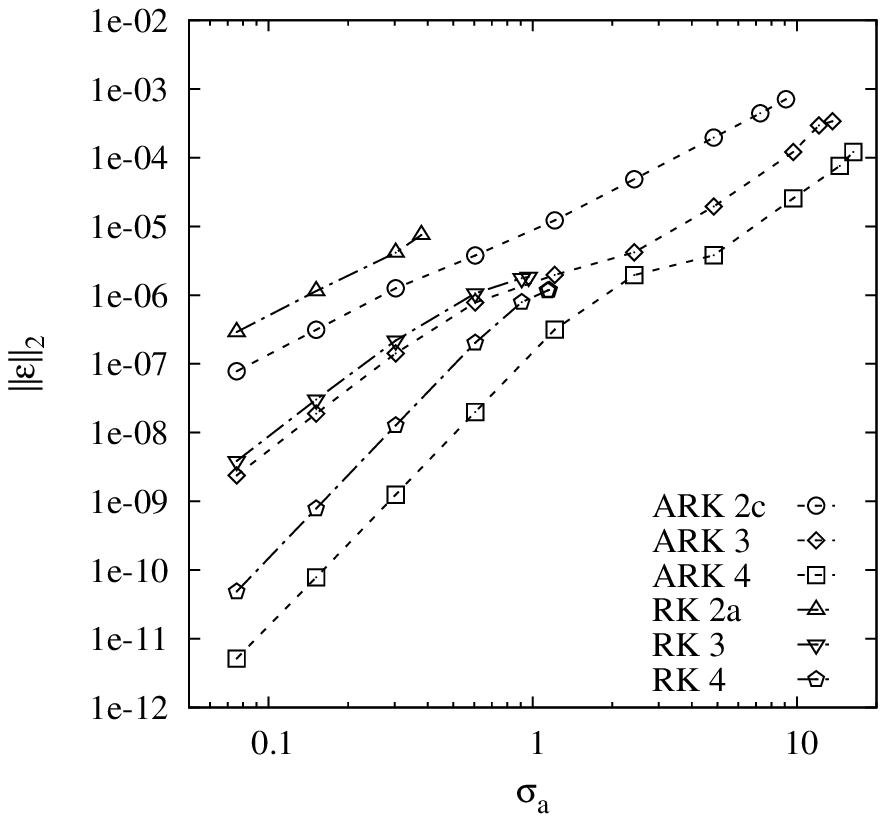}\label{fig:VortConv_ErrDt}}
\subfigure[Conservation error vs. acoustic CFL]{\includegraphics[width=0.49\textwidth]{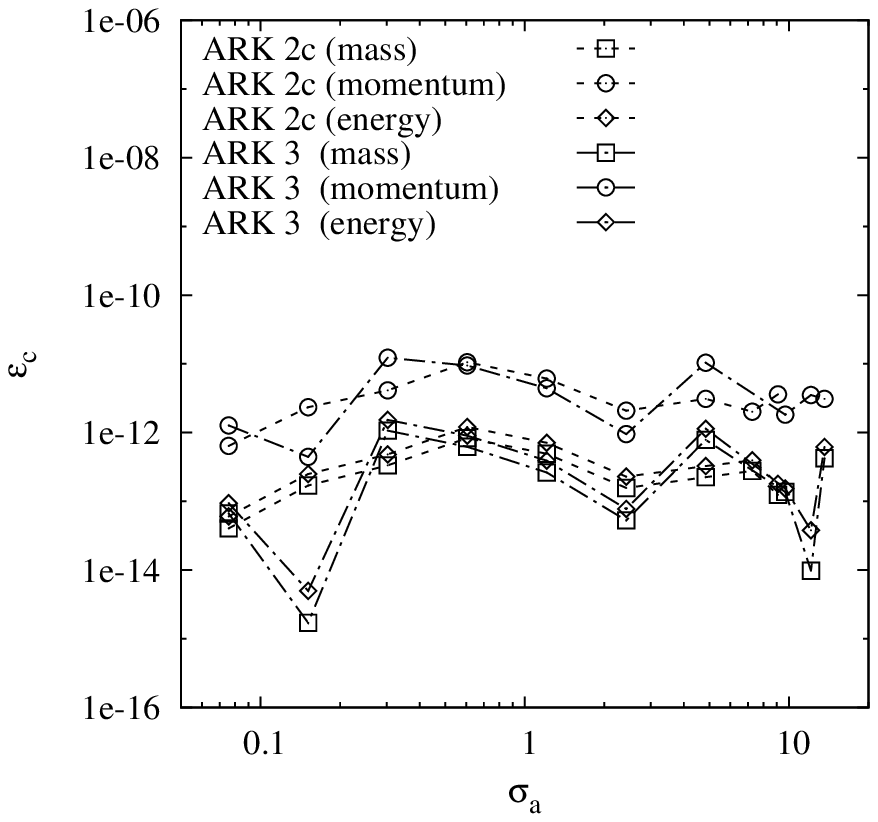}\label{fig:VortConv_ConsDt}}
\end{center}
\caption{Solution error ($\epsilon$) and conservation error ($\epsilon_c$) as a function of the acoustic CFL $\sigma_a$ for the isentropic vortex convection. The solutions are obtained on a $32^2$ grid with the WENO5 scheme at a final time of $100$ (one cycle over the domain).}
\label{fig:VortConv_Convergence}
\end{figure}

\begin{figure}[t!]
\begin{center}
\subfigure[ARK 2c, $\sigma_a \approx 7.6$]{\includegraphics[width=0.49\textwidth]{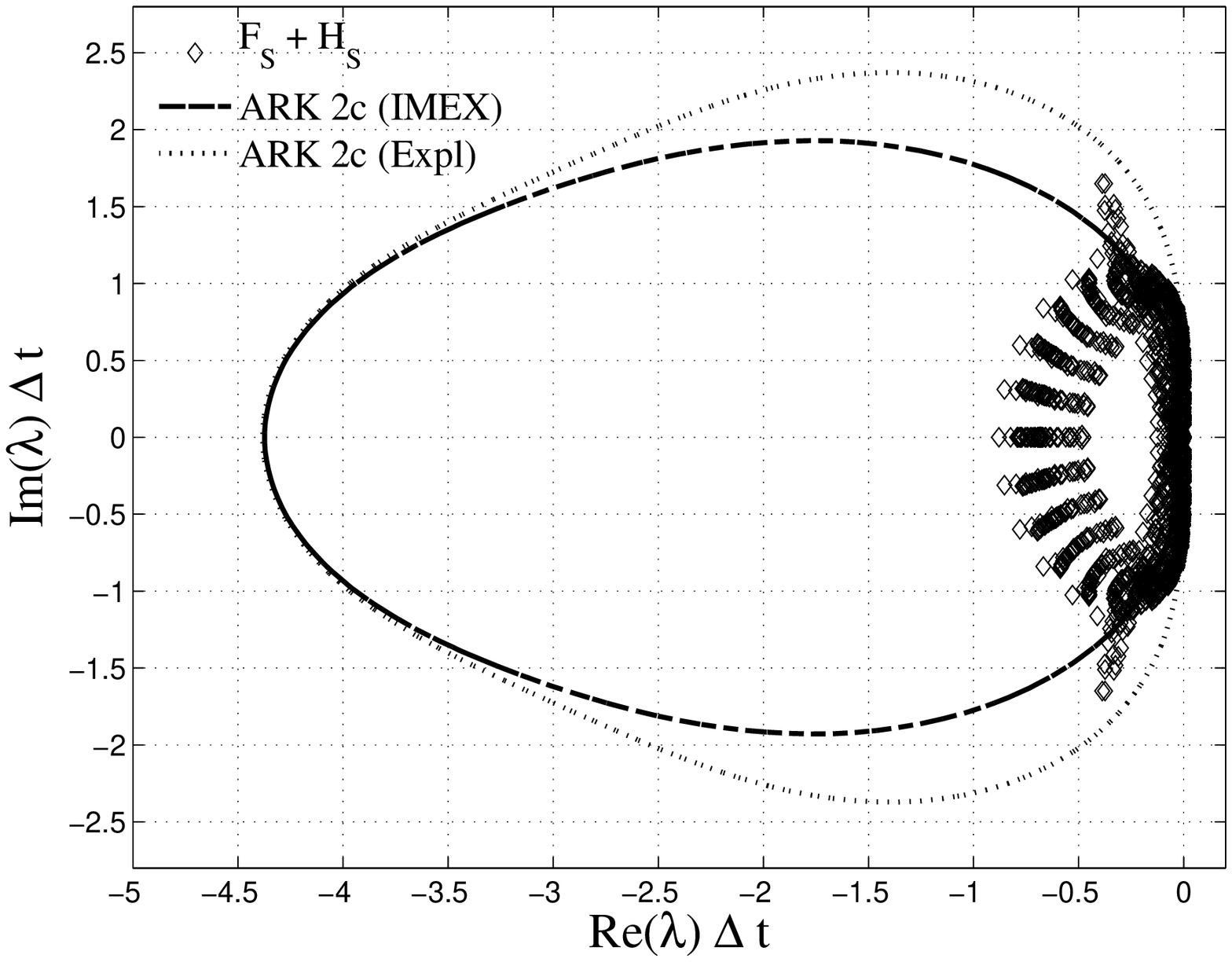}\label{fig:LowMachVortex_M0.10_WENO5_32_ARK2c_dt2.0}}
\subfigure[ARK 3,  $\sigma_a \approx 11.3$]{\includegraphics[width=0.49\textwidth]{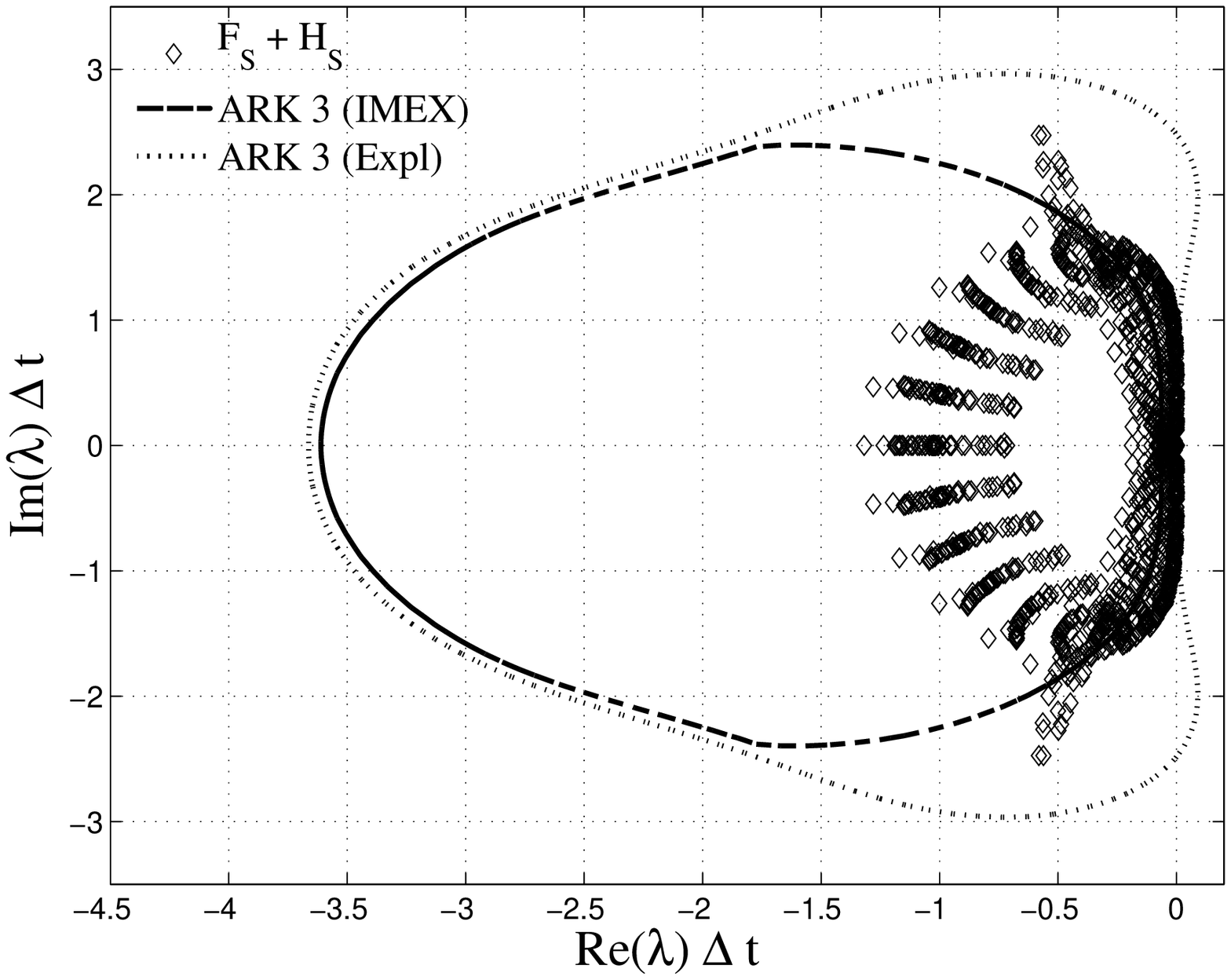} \label{fig:LowMachVortex_M0.10_WENO5_32_ARK3_dt3.0} }
\end{center}
\caption{Eigenvalues of the slow partitioned term $\left(\hat{\bf F}_S+\hat{\bf H}_S\right)$ multiplied by the time step $\Delta t$, and the stability regions of the explicit components of the ARK time integration methods. ``IMEX" denotes the stability region of the explicit method when the implicit method handles the eigenvalues of $\left(\hat{\bf F}_F+\hat{\bf H}_F\right)$, and ``Expl" denotes its stability region when it is used by itself as an explicit time integrator. }
\label{fig:VortConv_Stability}
\end{figure}

The error as a function of the acoustic CFL is shown in Figure~\ref{fig:VortConv_ErrDt}. The solutions are obtained on a grid with $32^2$ points with the WENO5 scheme after one cycle over the periodic domain. The tolerances for the GMRES solver are specified as $\tau_a = \tau_r = 10^{-10}$. We start the tests with an initially small time step and increase it until it reaches the stability limit of the time integrator being used. The error and the acoustic CFL are defined as
\begin{equation}\label{eqn:errorcfl}
{\bf \epsilon} = {\bf Q}\left(x,y,t\right) - {\bf Q}_{\rm ref}\left(x,y,t\right),\ \ \ \ \sigma_a = a_\infty \frac{\Delta t}{\min\left(\Delta x, \Delta y\right)},
\end{equation}
where ${\bf Q}_{\rm ref}\left(x,y,t\right)$ is the reference solution obtained with the explicit RK 4 time integration method with a very small time step of $0.0005$. The ARK methods converge at their theoretical orders for acoustic CFL numbers less than $1$; at higher CFL numbers, the acoustic mode is not resolved, and thus convergence is only second order. However, the absolute errors for a higher-order ARK method (say, ARK 4) are smaller than those for a lower-order ARK method (say, ARK 2c). The largest stable time step for the ARK methods are larger than those of the explicit RK methods by a factor of approximately $M_\infty^{-1}$, thus demonstrating that the time step size is determined by the advective scale. Figure~\ref{fig:VortConv_Stability} shows the eigenvalues of the slow operator  $\left(\hat{\bf F}_S+\hat{\bf H}_S\right)$ scaled by the time step $\Delta t$ and the stability regions of the explicit components of the ARK 2c and ARK 3 methods. The time step $\Delta t$ corresponds to acoustic CFL numbers of $\sim 7.6$ for ARK 2c and $\sim 11.3$ for ARK 3. These are close to the observed largest stable CFL numbers for these methods in Figure~\ref{fig:VortConv_ErrDt}. At these time step sizes, the advective eigenvalues have started spilling out of the respective stability regions. Comparison of the stability regions of the explicit method by itself (denoted by ``Expl") and when it is a part of an ARK method with the implicit part handling the eigenvalues of  $\left(\hat{\bf F}_F+\hat{\bf H}_F\right)$ (denoted by ``IMEX") shows significant reduction in the imaginary stability~\cite{constasandu2010}.

Figure~\ref{fig:VortConv_ConsDt} shows the conservation errors $\epsilon_c$ for mass ($\rho$), momentum ($\rho{\bf u}$), and energy ($e$) as a function of the acoustic CFL, for the ARK 2c and ARK 3 methods. The conservation error is defined as
\begin{equation}\label{eqn:conserror}
\epsilon_c = \frac{1}{\bar{Q}^k\left(0\right)} \left[\bar{Q}^k\left(t\right) - \bar{Q}^k\left(0\right) \right],\ \ \bar{Q}^k\left(t\right) = \int_V \|{\bf Q}^k\left(x,y,t\right)\|_2 dV,
\end{equation}
where $\bar{Q}$ is the volume integral over the domain, $V$ denotes the two-dimensional domain, and the superscript $k$ denotes the component ($k=1$ for mass, $k=2,3$ for momentum, and $k=4$ for energy). The conservation errors are on the order of round-off errors with the specified GMRES tolerances, for both the methods and at all the CFL numbers considered. In addition, they do not show any trends with respect to the CFL number. This demonstrates that the partitioned semi-implicit algorithm is conservative.

\section{Application to Atmospheric Flows}
\label{sec:atmosflows}

In this section, the algorithm is applied to atmospheric flows, which are governed by the two-dimensional Euler equations with a gravitational source term. Two benchmark flow problems are solved---the inertia-gravity wave and the rising thermal bubble. The flow solver used in this study has been previously verified for atmospheric flows with explicit Runge-Kutta schemes~\cite{ghoshconstaAIAAJ2016}; therefore, the focus of this section is to demonstrate the accuracy, stability, and numerical cost of the ARK methods. We note that the problems solved in this section are in terms of dimensional quantities, unlike the previous section where all quantities were nondimensional.

\subsection{Inertia-Gravity Waves}
\label{subsec:igwave}

\begin{figure}
\begin{center}
\includegraphics[width=0.93\textwidth]{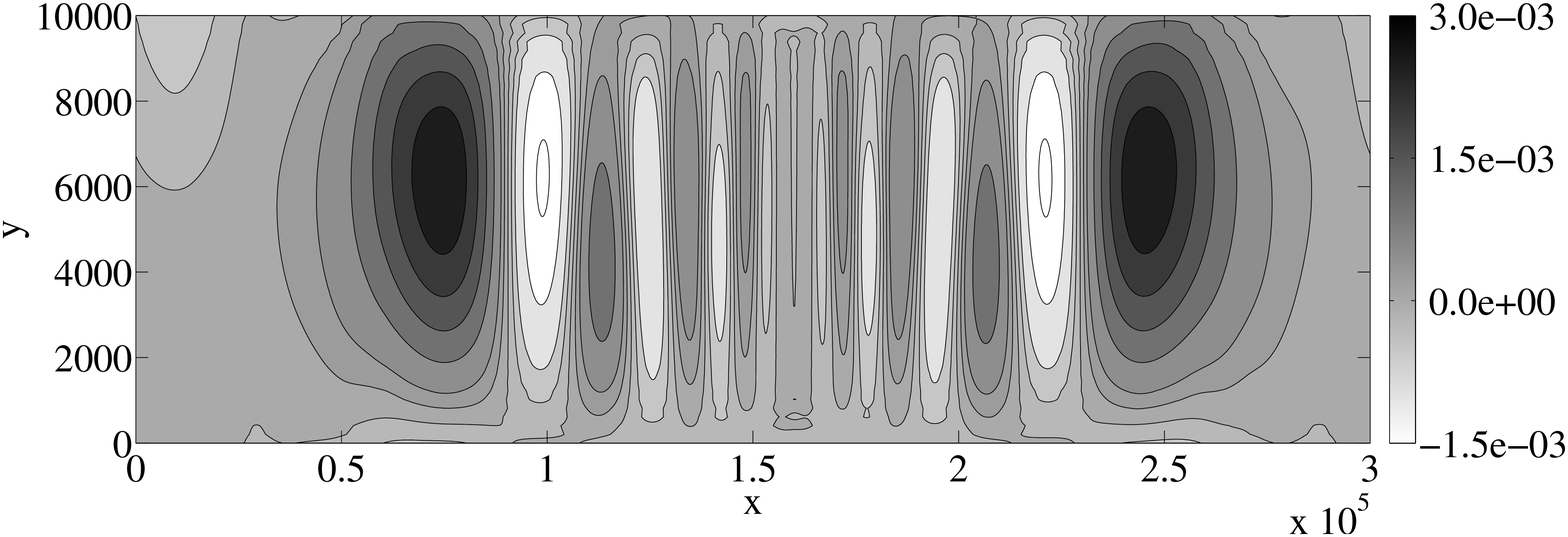}
\end{center}
\caption{Inertia-gravity waves: Potential temperature perturbation $\Delta \theta$ at $t = 3000\,\textup{s}$, obtained with the CRWENO5 scheme and the ARK 4 time integrator on a grid with $1200\times50$ points. The time step is $\Delta t = 12\,\textup{s}$, corresponding to an acoustic CFL number of $\sigma_a \approx 20.8$.}
\label{fig:igwave_contours}
\begin{center}
\includegraphics[width=0.93\textwidth]{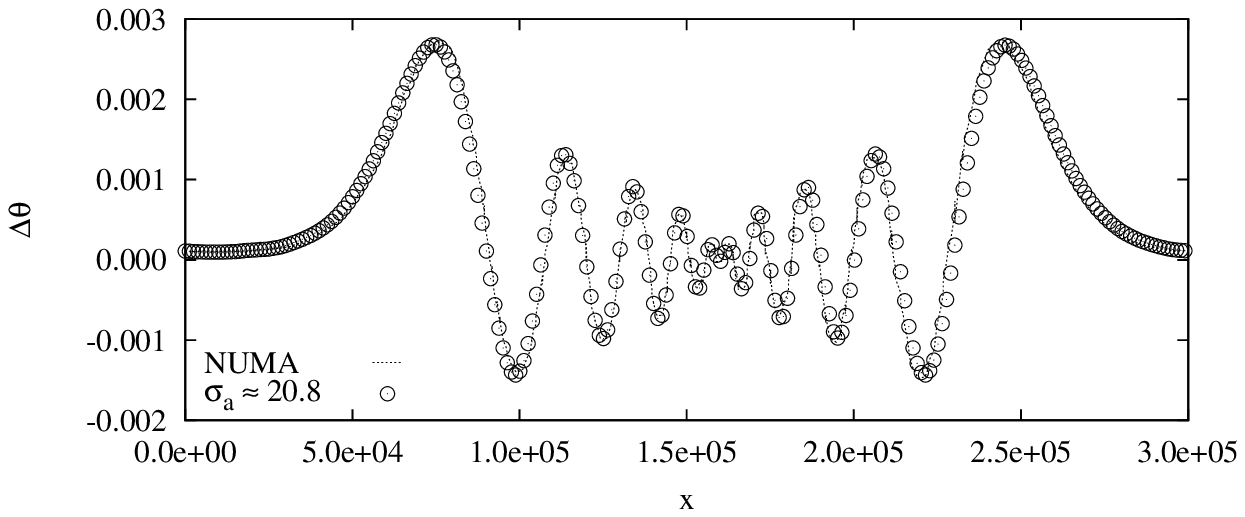}
\end{center}
\caption{Inertia-gravity waves: Cross-sectional potential temperature perturbation $\Delta \theta$ at $y = 5000\,\textup{m}$ and $t = 3000\,\textup{s}$, obtained with the CRWENO5 scheme and the ARK 4 time integrator on a grid with $1200\times50$ points. ``NUMA" refers to the reference solution obtained with a spectral element solver~\cite{giraldorestelli2008}.}
\label{fig:igwave_crosssec}
\end{figure}

\begin{figure}[h!]
\begin{center}
\subfigure[Eigenvalues]{\includegraphics[width=0.49\textwidth]{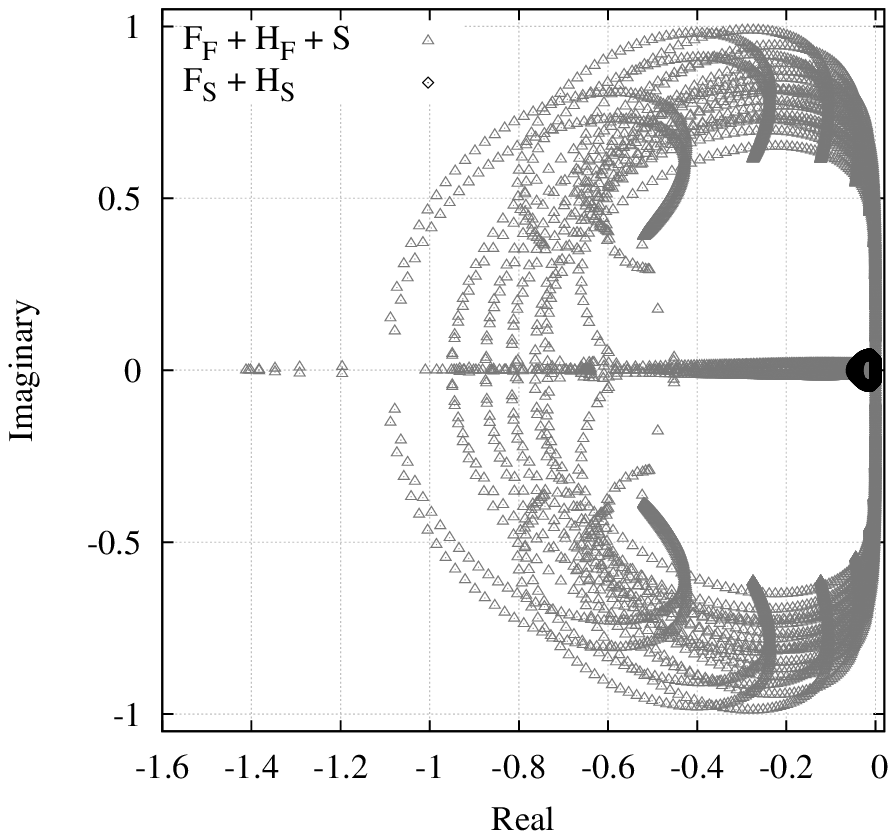}\label{fig:eigenvalues_igwave}}
\subfigure[Solution error as a function of acoustic CFL]{\includegraphics[width=0.49\textwidth]{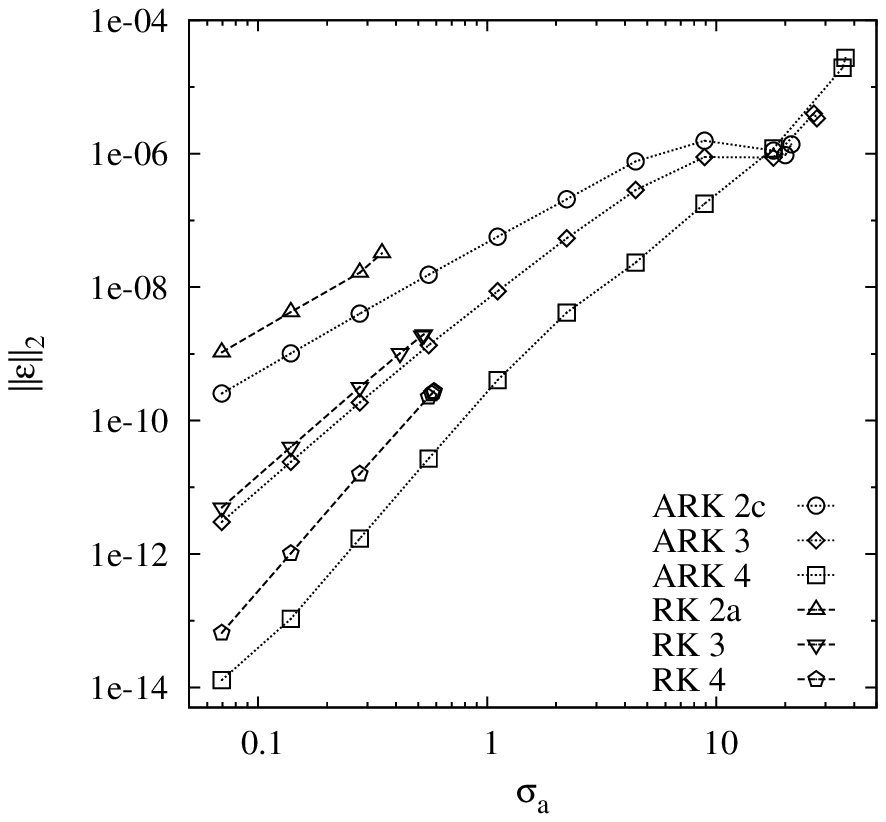}\label{fig:IGWave_ErrDt}}
\end{center}
\caption{Inertia-gravity waves: Eigenvalues of Jacobians of the partitioned terms $\left(\hat{\bf F}_F+\hat{\bf H}_F\right)+\hat{\bf S}$ and $\left(\hat{\bf F}_S+\hat{\bf H}_S\right)$ in (\ref{eqn:semidisc_ode_part_2d}), and the solution error ($\epsilon$) as a function of the acoustic CFL $\sigma_a$.}
\label{fig:IGWave_meow}
\end{figure}

The inertia-gravity wave~\cite{skamarockklemp,giraldorestelli2008} involves the evolution of a potential temperature perturbation. The domain is a channel with dimensions $300,000\,\textup{m} \times 10,000\,\textup{m}$. The initial flow consists of a perturbation introduced into a hydrostatically balanced (stratified) atmosphere. The mean flow is the stratified atmosphere with a specified Brunt-V{\"a}is{\"a}l{\"a} frequency ($\mathcal{N}$). The potential temperature and Exner pressure are given by
\begin{align}
\theta = T_0 \exp \left( \frac{\mathcal{N}^2}{g} y  \right), \pi = 1 + \frac{(\gamma-1)g^2}{\gamma R T_0 \mathcal{N}^2} \left[ \exp \left(- \frac{\mathcal{N}^2}{g} y  \right) -1 \right],
\end{align}
and the density and pressure are
\begin{align}
p &= p_0 \left[ 1 + \frac{(\gamma-1)g^2}{\gamma R T_0 \mathcal{N}^2} \left\{ \exp \left(- \frac{\mathcal{N}^2}{g} y  \right) -1 \right\} \right]^{\gamma/(\gamma-1)}, \\
\rho &= \rho_0 \exp\left(-\frac{\mathcal{N}^2}{g}y\right) \left[ 1 + \frac{(\gamma-1)g^2}{\gamma R T_0 \mathcal{N}^2} \left\{ \exp \left(- \frac{\mathcal{N}^2}{g} y  \right) -1 \right\} \right]^{1/(\gamma-1)}.
\end{align}
The initial velocity components are $u=20\,\textup{m}/\textup{s}$ and $v = 0\,\textup{m}/\textup{s}$. Periodic boundary conditions are applied on the left ($x=0\,\textup{m}$) and right ($x=300,000\,\textup{m}$) boundaries, while inviscid wall boundary conditions are applied on the bottom ($y=0\,\textup{m}$) and top ($y=10,000\,\textup{m}$) boundaries. The Brunt-V{\"a}is{\"a}l{\"a} frequency is $\mathcal{N} = 0.01\,/\textup{s}$, and the gravitational force per unit mass is $9.8\,\textup{m}/\textup{s}^2$ along the $y$-direction. The reference pressure ($p_0$) and temperature ($T_0$) at $y=0\,\textup{m}$ are $10^5\,\textup{N}/\textup{m}^2$ and $300\,\textup{K}$, respectively, and the reference density is computed from the equation of state $p_0 = \rho_0 R T_0$. The universal gas constant $R$ is $287.058\,\textup{J}/\textup{kg K}$. The perturbation is added to the potential temperature, specified as
\begin{equation}
\Delta\theta\left(x,y,t=0\right) = \theta_c \sin \left( \frac{\pi_c y}{h_c} \right) \left[ 1 + \left( \frac{x-x_c}{a_c} \right)^2 \right]^{-1},
\end{equation}
where $\theta_c = 0.01\,\textup{K}$ is the perturbation strength, $h_c = 10,000\,\textup{m}$ is the height of the domain, $a_c = 5,000\,\textup{m}$ is the perturbation half-width, $x_c = 100,000\,\textup{m}$ is the horizontal location of the perturbation, and $\pi_c \approx 3.141592654$ is the Archimedes (trigonometric) constant. The evolution of the perturbation is simulated until a final time of $3000\,\textup{s}$. The reference speed of sound is $a_0 = \sqrt{\gamma R T_0} = 347.22\,\textup{m}/\textup{s}$, and the reference Mach number for this flow is approximately $0.06$. Figure~\ref{fig:eigenvalues_igwave} shows the eigenvalues of the slow and the fast operators for the problem discretized on a $300\times10$-point grid with the CRWENO5 scheme; the fast operator includes the gravitational source term.

Figure~\ref{fig:igwave_contours} shows the potential temperature perturbation $\Delta \theta = \left(\theta-\theta_0\right)$ contours for the solution obtained with the CRWENO5 scheme on a grid with $1200\times50$ points. The ARK 4 method is used for time integration with a time step of $\Delta t = 12\,\textup{s}$, corresponding to an acoustic CFL number of $\sigma_a \approx 20.8$. The tolerances for the GMRES solver are specified as $\tau_a = \tau_r = 10^{-6}$. A good agreement is observed with results in the literature~\cite{ahmadlindeman,skamarockklemp,ahmadproctor2012,yangcai2015}. Figure~\ref{fig:igwave_crosssec} shows the cross-sectional potential temperature perturbation at an altitude of $y=5,000\,\textup{m}$ for this solution. The reference solution ``NUMA" refers to the solution obtained with a spectral-element solver~\cite{giraldorestelli2008}, with $10$th-order polynomials, $3$rd-order explicit RK time integration, and $250\,\textup{m}$ effective grid resolution. The solution obtained with the partitioned semi-implicit approach agree well with the reference solution.

Figure~\ref{fig:IGWave_ErrDt} shows the $L_2$ norm of the solution error as a function of the acoustic CFL for solutions obtained with the CRWENO5 scheme on a $600\times20$ grid. 
The error and the acoustic CFL are as defined in (\ref{eqn:errorcfl}). The reference speed of sound $a_0$ is used to compute the acoustic CFL, and the reference solution is obtained with the explicit RK 4 time integrator and a very small time step of $0.005$. 
The tolerances for the GMRES solver are specified as $\tau_a = \tau_r = 10^{-10}$. 
The errors for the partitioned ARK methods are shown, as well as the explicit RK 2a, RK 3, and RK 4 methods. The ARK methods converge at their theoretical orders of convergence, and the largest stable time steps are observed to be approximately $M_\infty^{-1} \approx 15$ times larger than those of the explicit RK methods. The mass conservation errors are on the order of round-off errors for all the solutions and at all CFL numbers considered.

\subsection{Rising Thermal Bubble}
\label{subsec:rtb}

The two-dimensional rising thermal bubble~\cite{giraldorestelli2008} simulates the dynamics of a warm bubble. A square domain of size $1000\,\textup{m}\times1000\,\textup{m}$ is specified with inviscid wall boundary conditions on all sides. The initial solution is a warm bubble introduced in a hydrostatically balanced atmosphere. The mean flow is the stratified atmosphere with a constant potential temperature $\theta = T_0 = 300\,\textup{K}$; and the density, pressure, and velocity are respectively
\begin{equation}
\rho = \rho_0 \left[ 1 - \frac{(\gamma-1)gy}{\gamma R \theta} \right] ^ {1/(\gamma-1)},\ 
p    = p_0    \left[ 1 - \frac{(\gamma-1)gy}{\gamma R \theta} \right] ^ {\gamma/(\gamma-1)},\ u = v = 0.
\end{equation} 
The reference pressure is $10^5\,\textup{N}/\textup{m}^2$, and the reference density is computed from the equation of state $p_0 = \rho_0 R T_0$. The universal gas constant $R$ is $287.058\,\textup{J}/\textup{kg K}$. A constant gravitation force per unit mass of $9.8\,\textup{m}/\textup{s}^2$ is specified along the $y$-direction. The warm bubble is added as a potential temperature perturbation,
\begin{equation}
\Delta\theta\left(x,y,t=0\right) = \left\{ \begin{array}{cc} 0 & r > r_c \\ \frac{\theta_c}{2}\left[ 1 + \cos\left( \frac{\pi_c r}{r_c} \right) \right] & r \leq r_c \end{array} \right.,\ r = \sqrt{(x-x_c)^2+(y-y_c)^2},
\end{equation}
where $\theta_c = 0.5\,\textup{K}$ is the perturbation strength, $\left(x_c,y_c\right) = (500,350)\,\textup{m}$ is the initial location at which the bubble is centered, $r_c=250\,\textup{m}$ is the radius of the bubble, and $\pi_c$ is the trigonometric constant. The flow is simulated to a final time of $400\,\textup{s}$.

\begin{figure}[t!]
\begin{center}
\subfigure[$t = 0\,\textup{s}$  ]{\includegraphics[width=0.49\textwidth]{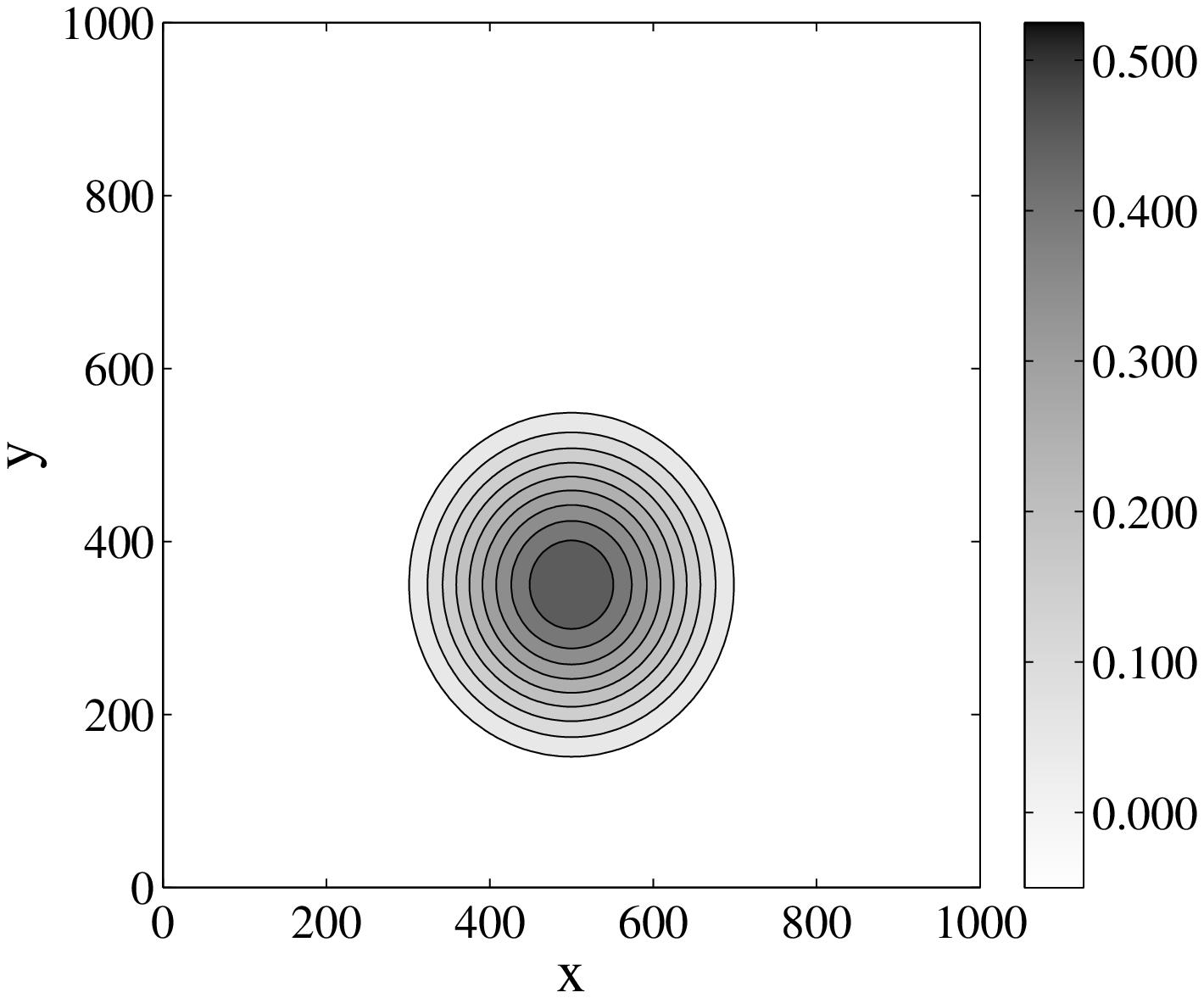}\label{fig:rtb_contour_000}}
\subfigure[$t = 400\,\textup{s}$]{\includegraphics[width=0.49\textwidth]{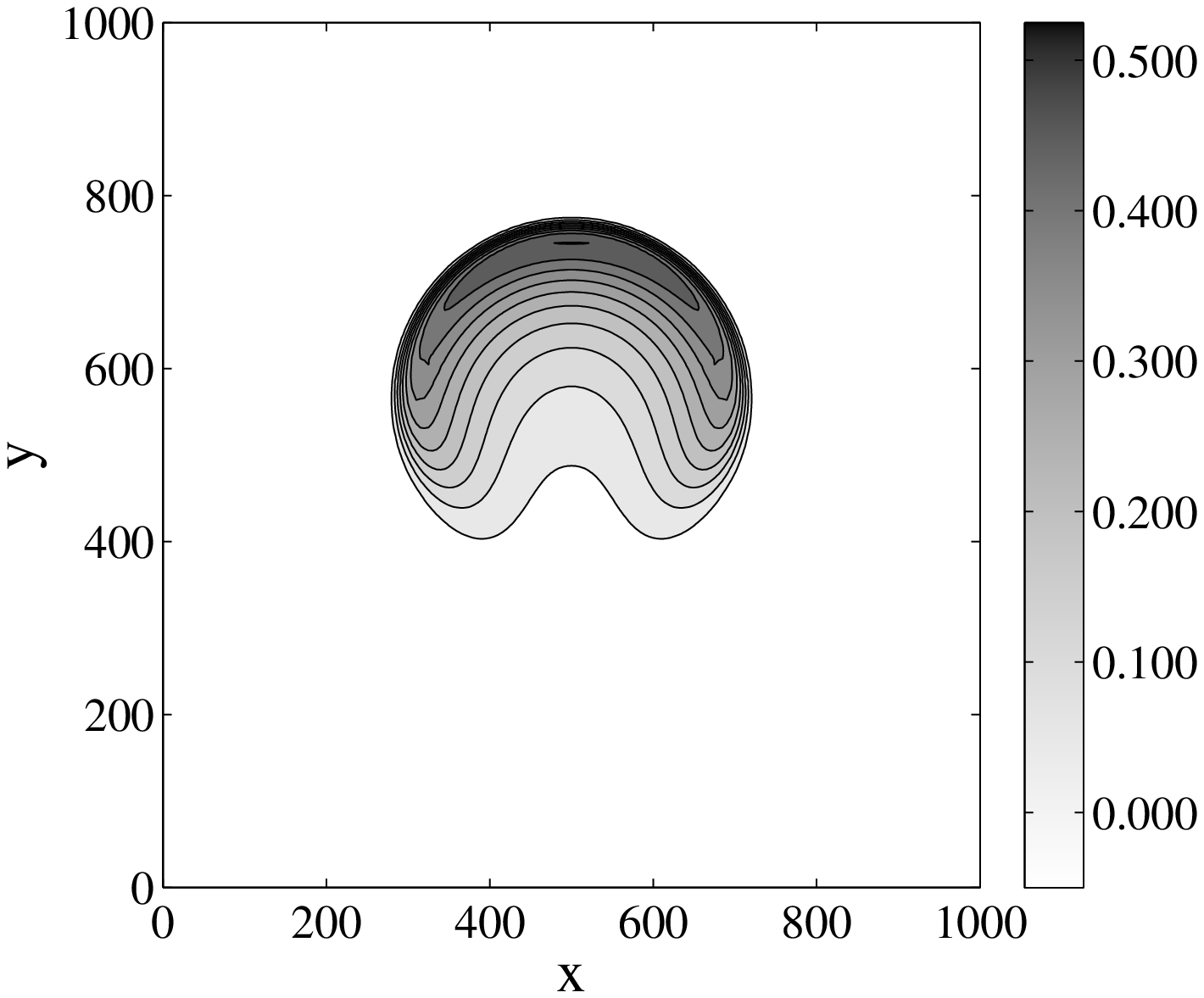}\label{fig:rtb_contour_400}}
\end{center}
\caption{Rising thermal bubble: Potential temperature perturbation $\Delta \theta$ for the solution obtained with the WENO5 scheme and the ARK 4 time integrator on a grid with $201^2$ points. The time step is $\Delta t = 2\,\textup{s}$, corresponding to an acoustic CFL number of $\sigma_a \approx 139$.}
\label{fig:rtb_contour}
\end{figure}


\begin{figure}[t!]
\begin{center}
\subfigure[Cross-sectional potential temperature perturbation at $y=550\,\textup{m}$]{\includegraphics[width=0.49\textwidth]{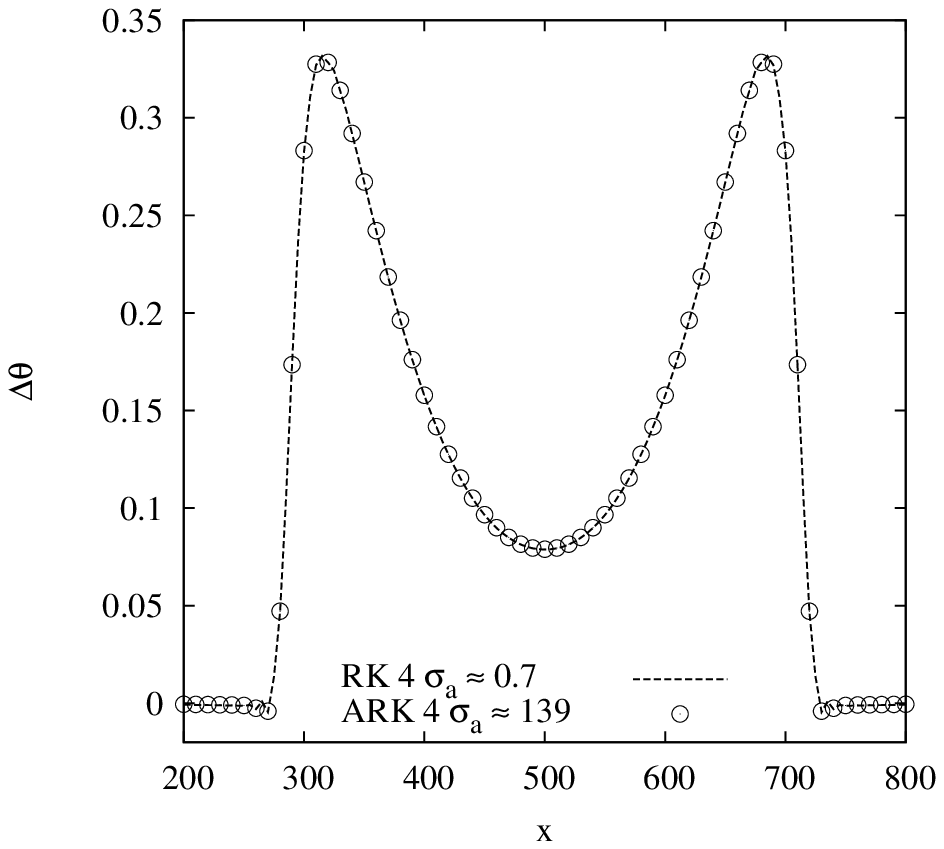}\label{fig:rtb_crosssec_x}}
\subfigure[Error vs. acoustic CFL]{\includegraphics[width=0.49\textwidth]{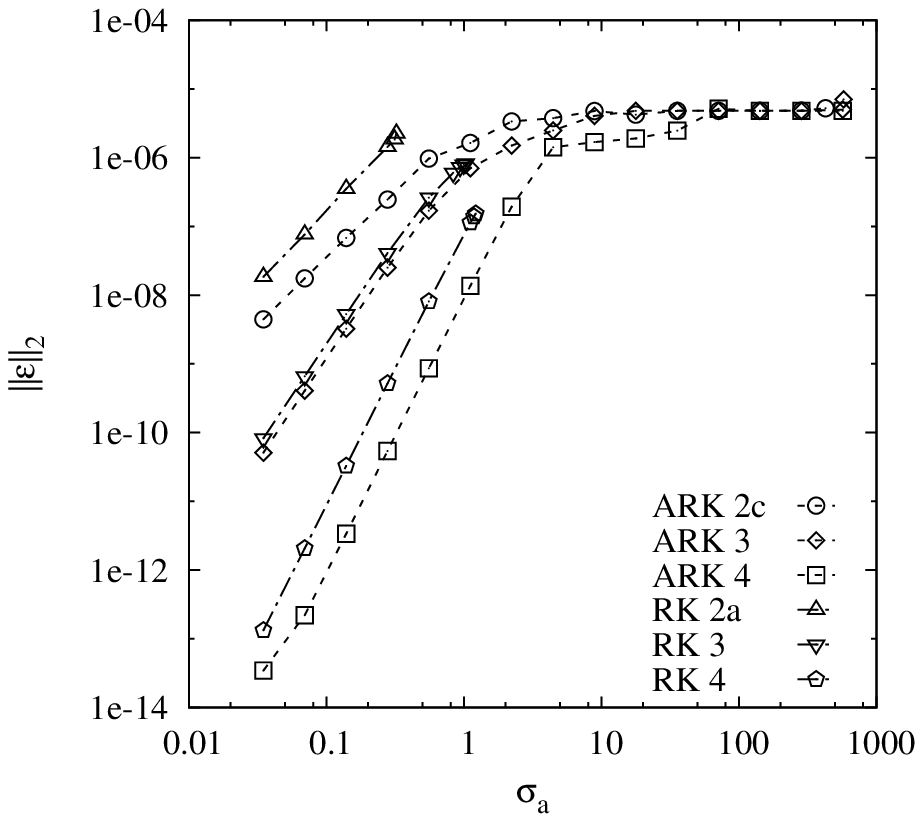}\label{fig:RTB_ErrDt}}
\end{center}
\caption{Rising thermal bubble: Comparison of cross-sectional solution profile ($201^2$-points grid), and solution error ($\epsilon$) as a function of the acoustic CFL $\sigma_a$ ($51^2$-points grid) at a final time of $400\,\textup{s}$.}
\end{figure}

Figure~\ref{fig:rtb_contour} shows the initial solution at $t=0\,\textup{s}$ and the solution at $t=400\,\textup{s}$. The warm bubble rises as a result of buoyancy and deforms as a result of the temperature and velocity gradients. The potential temperature perturbation $\Delta \theta = \theta - \theta_0$ is shown. The solution is obtained with the WENO5 scheme and the ARK 4 time integrator on a grid with $201^2$ points. The GMRES solver tolerances are $\tau_a = \tau_r = 10^{-6}$. The time step size is $2\,\textup{s}$, which results in an acoustic CFL number of approximately $139$. The acoustic CFL is given by (\ref{eqn:errorcfl}), and the reference speed of sound $a_0$ is used. The flow is initially at rest, and thus the advective eigenvalues are all zero. As the bubble rises, it induces a velocity field; at $t=400\,\textup{s}$, the maximum velocity magnitude in the domain is approximately $2.1\,\textup{m}/\textup{s}$, corresponding to a maximum local Mach number of approximately $0.006$. Thus, the disparity between the advective and acoustic scales is very large, and the semi-implicit approach allows time steps that are much larger than those allowed by an explicit time integrator. Figure~\ref{fig:rtb_crosssec_x} compares the cross-sectional profiles of $\Delta \theta$ along $x$ at $y=550\,\textup{m}$ for this solution and that obtained with the explicit RK 4 method at an acoustic CFL number of $\sim0.7$, and an excellent agreement is observed.

The $L_2$ norm of the solution error is shown in Figure~\ref{fig:RTB_ErrDt} as a function of the acoustic CFL. The error, defined in (\ref{eqn:errorcfl}), is computed with respect to a reference solution that is obtained on the same grid with the same spatial discretization and with the explicit RK 4 time integrator with a very small time step size of $10^{-4}$. The figure shows the errors for the ARK 2c, ARK 3, and ARK 4 methods, as well as the explicit RK 2a, RK 3, and RK 4 methods. The tolerances specified for the GMRES solver are $\tau_a = \tau_r = 10^{-10}$. In the region where the acoustic waves are resolved and the explicit methods are stable, all the methods converge at their theoretical orders of accuracy. At higher CFL numbers, the acoustic mode is not resolved, and thus the errors for the ARK methods (relative to the reference solution, in which the acoustic mode is resolved) converge toward a similar value. This behavior has been previously analyzed and discussed for the semi-implicit time integration of the perturbation form of the governing equations~\cite{giraldokellyconsta2013}. The mass conservation error are zero to machine precision for all the solutions at all the CFL numbers considered.

\subsection{Numerical Cost}
\label{subsec:numcost}

The main objective of using semi-implicit time integration is to obtain well-resolved solutions at a lower computational cost than with explicit time integrators. These methods allow time step sizes that step over the fast acoustic scales; however, they require the solution of a system of equations. Thus, their performance depends on the cost and accuracy of the linear solver. In this section, we compare the computational cost of the ARK methods with the explicit RK methods in terms of the minimum wall time and the number of function calls required to obtain a stable and resolved solution. In the following discussion, the number of function calls ($n_{\rm FC}$) refers to the total number of calls to the functions that compute the partitioned flux components $\hat{\bf F}_F$ or $\hat{\bf F}_S$. Since a matrix-free implementation of the Jacobian is used, $n_{\rm FC}$ is the sum of the total number of time iterations ($n_T$) times the number of stages $s$ (of the time integration method), and the total number of GMRES iterations. It is thus an estimate of the total computational cost; however, it does not include the cost of assembling and inverting the preconditioning matrix. The algorithm is implemented in serial; its performance and scalability on parallel platforms are being investigated. The reported simulations are run on a $2200\,\textup{MHz}$ AMD Opteron processor.

\begin{table}[t!]
\begin{center}
\caption{Inertia-gravity waves: $L_2$ norm of the error and computational cost as a function of time step size and acoustic CFL number of the ARK and RK methods for solutions on a grid with $1200\times50$ points discretized in space with the CRWENO5 scheme. The final time is $3000\,\textup{s}$. Boldfaced rows indicate the performance at the largest stable time step for the ARK methods.}
\label{tab:igwave}
\begin{tabular*}{\textwidth}{@{\extracolsep{\fill}}|l | r | r | r | r | r | r |}
\hline
Method & $\Delta t$ & $\|\epsilon\|_2$ & $n_{\rm T}$ & $\sigma_a$ & $n_{\rm FC}$ & Wall time (s) \\
\hline
RK 2a & $0.15$ & $1.4\times10^{-8}$ & $20,000$ & $0.26$ & $40,000$ & $12,353$ \\
RK 4  & $0.30$ & $1.7\times10^{-9}$ & $10,000$ & $0.45$ & $40,000$ & $12,072$ \\
\hline
ARK 2c & $2.0$ & $4.6\times10^{-7}$ & $1,500$ & $3.47$  & $57,121$ & $23,180$\\
       & $4.0$ & $1.3\times10^{-6}$ & $750$  & $6.94$  & $34,530$ & $14,086$\\
       & $\bf8.0$ & ${\bf 9.1\times10^{-7}}$ & $\bf375$  & $\bf13.89$ & $\bf21,164$ & $\bf8,797$ \\
\hline
ARK 4  & $4.0$  & $1.9\times10^{-8}$ & $750$  & $6.94$  & $80,479$  & $33,296$\\
       & $8.0$  & $2.0\times10^{-7}$ & $375$  & $13.89$ & $47,258$  & $19,875$\\
       & $12.0$ & $5.1\times10^{-7}$ & $250$  & $20.83$ & $34,900$  & $14,398$\\
       & $\bf15.0$ & ${\bf9.2\times10^{-7}}$ & $\bf200$  & $\bf26.04$ & $\bf29,556$  & $\bf12,608$\\
\hline
\end{tabular*}
\end{center}
\end{table}

Table~\ref{tab:igwave} shows the wall times (in seconds), the number of function calls, and the $L_2$ norm of the error $\epsilon$ for the inertia-gravity wave problem, solved on a grid with $1200\times50$ points with the CRWENO5 scheme. The tolerances for the GMRES solver are $\tau_a = \tau_r = 10^{-6}$. The ARK 2c and ARK 4 methods are compared with the explicit RK 2a and RK 4. The time steps for the explicit RK methods are chosen close to their stability limits; thus, the reported wall times are the fastest time to solution for the explicit methods. The final row for each ARK method reports the cost with the largest stable time step and thus represents their fastest time to solution. The cost of the ARK methods decreases as the time step size increases (both the number of function calls and the wall times). ARK 2c is the fastest method among those considered. The acoustic scale is approximately $17$ times faster than the advective scale for this problem. While the ARK 2c is faster than the explicit methods by $~25\%$, the ARK 4 is generally slower at all the CFL numbers except at the largest CFL, where its cost is comparable. The solution errors are consistent with those reported in Figure~\ref{fig:IGWave_ErrDt}, and thus the larger tolerances for the GMRES solver used for the performance tests ($\tau_{a,r}$) do not degrade the accuracy of the overall solution. Since we are considering large time step sizes, a more relaxed tolerance suffices to ensure that the error in solving the implicit stages remains small with respect to the truncation error of the time integration scheme.

\begin{table}[t!]
\begin{center}
\caption{Rising thermal bubble: $L_2$ norm of the error and computational cost as a function of time step size and acoustic CFL number of the fourth-order ARK and RK methods for solutions on a grid with $201^2$ points discretized in space with the WENO5 scheme. The final time is $400\,\textup{s}$. Boldfaced rows indicate the performance at the largest stable time step for the ARK method.}
\label{tab:rtb}
\begin{tabular*}{\textwidth}{@{\extracolsep{\fill}}|l | r | r | r | r | r | r |}
\hline
Method & $\Delta t$ & $\|\epsilon\|_2$ & $n_{\rm T}$ & $\sigma_a$ & $n_{\rm FC}$ & Wall time (s) \\
\hline
RK 4  &  $0.01$ & $7.5\times10^{-8}$ & $40,000$ & $0.69$ & $160,000$ & $30,154$\\
\hline
ARK 4  & $0.10$ & $1.5\times10^{-7}$ & $4,000$  & $6.94$   & $360,016$  & $73,111$  \\
       & $0.50$ & $1.6\times10^{-6}$ & $800$   & $34.72$  & $111,824$  & $22,104$  \\
       & $\bf 2.00$ & $\bf1.9\times10^{-6}$ & $\bf 200$   & $\bf 138.89$ & $\bf 45,969$   & $\bf 8,569$   \\
\hline
\end{tabular*}
\end{center}
\end{table}

Table~\ref{tab:rtb} shows the cost of the ARK 4 method and the $L_2$ norm of the numerical error for the rising thermal bubble, solved on a grid with $201^2$ points with the WENO5 scheme. The tolerances for the GMRES solver are $\tau_a = \tau_r = 10^{-6}$. The cost of the explicit RK 4 method is used as a reference. The separation between the acoustic and advective scales is very large; the flow is initially at rest, with the Mach number at the final time being $\sim0.006$. The semi-implicit method is thus able to take much larger time steps. The cost of the ARK method decreases as the time step size increases; and for CFL numbers greater than $\sim30$, the ARK 4 is faster than the RK 4. At the largest stable time step, the ARK 4 is faster than the RK 4 method by a factor of approximately $3.5$. The numerical errors are consistent with those reported in Figure~\ref{fig:RTB_ErrDt} thus ensuring that the relaxed tolerance for the GMRES solver does not compromise the accuracy of the time integration.

The results reported here are obtained with basic preconditioning of the linear system, as described in Section~\ref{subsec:precon}. The primary focus of this paper is to introduce a flux partitioning for semi-implicit time integration based on the governing equations expressed as (\ref{eqn:mass_cons})--(\ref{eqn:energy_cons}). Improving the efficiency of the time integrator by developing suitable preconditioning techniques for the GMRES solver is currently being investigated.

\section{Conclusion}
\label{sec:conc}

This paper presents a characteristic-based partitioning of the
hyperbolic flux in the compressible Euler equations for semi-implicit
time integration. The acoustic and the advective modes are separated;
the former is integrated in time implicitly because of its stiffness,
while the latter is integrated explicitly. The stiff term is
linearized, and thus the semi-implicit algorithm requires only the
solution to a linear system. The nonstiff term, defined as the total
nonlinear flux with the linearized stiff term subtracted from it, is
treated explicitly. High-order additive Runge-Kutta methods are
applied to the partitioned equations, and the WENO and CRWENO schemes
are used for the spatial discretization. We note that Rosenbrock
schemes are also viable time stepping alternatives.

We test this approach on simple inviscid flow problems at low Mach numbers. The results show that the largest stable time step is determined by the advective scale. The algorithm is then applied to atmospheric flows where the acoustic modes are much faster than the advective mode but are not physically relevant. The accuracy and convergence of the algorithm are demonstrated for benchmark problems, and the results show that the partitioned semi-implicit approach is conservative. Moreover, the computational cost is assessed and compared with that of explicit time integrators. The extension of this algorithm to parallel platforms and the development of more effective preconditioning techniques are areas of current research.

\bibliographystyle{siam}
\bibliography{atmos,crweno,time}
\end{document}